%% file: decentralized_RA_v10_journal.tex
\documentclass[journal, 12pt, onecolumn,draftcls]{IEEEtran}
\usepackage{amsmath}
\usepackage{amssymb}
\usepackage{graphics,graphicx,multicol}
\usepackage{color}
\usepackage{epsfig}
  
\newcommand{\be}{\begin{eqnarray}}
\newcommand{\ee}{\end{eqnarray}}
\newcommand{\beq}{\begin{eqnarray}}
\newcommand{\eeq}{\end{eqnarray}}
\newcommand{\n}{\nonumber}
\newcommand{\C}{\mathcal{C}}
\newcommand{\X}{\mathcal{X}}
\newcommand{\R}{\mathcal{R}}
\newcommand{\T}{\mathcal{T}}
\newcommand{\calS}{\mathcal{S}}
\newcommand{\calL}{\mathcal{L}}
\newcommand{\bbR}{\mathbb{R}}
\newcommand{\bbZ}{\mathbb{Z}}
\newcommand{\bbE}{\mathbb{E}}
\newcommand{\bc}{\mathbf{c}}
\newcommand{\br}{\mathbf{r}}
\newcommand{\bQ}{\mathbf{Q}}
\newcommand{\bI}{\mathbf{I}}
\newcommand{\ba}{\mathbf{a}}
\newcommand{\bb}{\mathbf{b}}
\newcommand{\bt}{\mathbf{t}}
\newcommand{\bv}{\mathbf{v}}
\newcommand{\bde}{\mathbf{e}}
\newcommand{\bx}{\mathbf{x}}
\newcommand{\bs}{\mathbf{s}}
\newcommand{\ei}{\mathbf{e}_i}
\newcommand{\blam}{\boldsymbol\lambda}
\newcommand{\bet}{\boldsymbol\eta}
\newcommand{\btheta}{\boldsymbol\theta}
\newcommand{\bmu}{\boldsymbol\mu}
\newcommand{\bpi}{\boldsymbol\pi}
\newcommand{\balpha}{\boldsymbol\alpha}
\newcommand{\bbeta}{\boldsymbol\beta}

\newcommand{\bsig}{\boldsymbol\sigma}

\newcommand{\cL}{{\mathcal L}}
\newcommand{\cN}{\mathcal{N}}
\newcommand{\cB}{\mathcal{B}}
\newcommand{\cG}{{\mathcal G}}

\newcommand{\cE}{{\mathcal E}}

\newcommand{\cS}{{\mathcal S}}

\begin{document}

\newtheorem{theorem}{Theorem}
\newtheorem{lemma}[theorem]{Lemma}
\newtheorem{remark}{Remark}
\newtheorem{result}{Result}
\newtheorem{defn}{Definition}
\newtheorem{exm}{Example}

\title{Distributed Rate Allocation for Wireless Networks}
\author{Jubin~Jose
        and~Sriram~Vishwanath
\thanks{J. Jose and S. Vishwanath are with the Department of Electrical and Computer Engineering, The University of Texas at Austin, Austin, TX 78712 USA (email(s): jubin@austin.utexas.edu; sriram@austin.utexas.edu).}
}
\maketitle
\begin{abstract}
This paper develops a distributed algorithm for rate allocation in wireless networks that achieves the same throughput region as optimal centralized algorithms. This cross-layer algorithm jointly performs medium access control (MAC) and physical-layer rate adaptation. The paper establishes that this algorithm is throughput-optimal for general rate regions. In contrast to on-off scheduling, rate allocation enables optimal utilization of physical-layer schemes by scheduling multiple rate levels. The algorithm is based on local queue-length information, and thus the algorithm is of significant practical value.

The algorithm requires that each link can determine the global feasibility of increasing its current data-rate. In many classes of networks, any one link's data-rate primarily impacts its neighbors and this impact decays with distance. Hence, local exchanges can provide the information needed to determine feasibility. Along these lines, the paper discusses the potential use of existing physical-layer control messages to determine feasibility. This can be considered as a technique analogous to carrier sensing in CSMA (Carrier Sense Multiple Access) networks. An important application of this algorithm is in multiple-band multiple-radio throughput-optimal distributed scheduling for white-space networks.
\end{abstract}



\section{Introduction}
\label{sec-introduction}

The throughput of wireless networks is traditionally studied separately at the physical and medium access layers, and thus independently optimized at each of these two layers. As a result, conventionally, data-rate adaptation is performed at the physical layer for each link, and link scheduling is performed at the medium access layer. There are significant throughput gains in studying these two in a cross-layer framework \cite{J:TE92,Eryilmaz-Srikant-Perkins-05,B:GNT06,Lin_Shroff_Srikant_06,Chiang_Low_Calderbank_Doyle_07}. This cross-layer optimization results in a joint rate allocation for all the links in the network.

Maximum Weighted (Max-Weight) scheduling introduced in the seminal paper \cite{J:TE92} performs joint rate allocation and guarantees throughput-optimality\footnote{For cooperative networks, throughput-optimal rate allocation does not follow from classical Max-Weight scheduling. In \cite{Jose-Ying-Vishwanath-09}, modified algorithms are developed for certain cooperative networks that guarantee throughput-optimality.}. However, Max-Weight algorithm and its variants have the following disadvantages. (\emph{a}) It requires periodic solving of a possibly hard optimization problem. (\emph{b}) The optimization problem is centralized, and thus introduces significant overhead due to queue-length information exchanges. Thus, in order to overcome these disadvantages, we need efficient distributed algorithms for general physical-layer interference models \cite{Lin_Shroff_Srikant_06}.

The goal of this paper is to perform joint rate allocation in a decentralized manner. A related problem is distributed resource allocation in networks, and this problem has received considerable attention in diverse communities over years. In data and/or stochastic processing networks, resource-sharing is typically described in terms of independent set constraints. With such independent set constraints, the resource allocation problem translates to medium access control (or link scheduling) in wireless networks. For such on-off scheduling, recently, efficient algorithms have been proposed for both random access networks \cite{Gupta_Stolyar_06,Stolyar-08} and CSMA networks \cite{Marbach_Eryilmaz_Ozdaglar_07,Bordenave-08}. More recently, with instantaneous carrier sensing, a throughput-optimal algorithm with local exchange of control messages that approximate Max-Weight has been proposed in \cite{Rajagopalan-09}, and a fully decentralized algorithm has been proposed in \cite{Jiang-Allerton}. The decentralized queue-length based scheduling algorithm in \cite{Jiang-Allerton} and its variants have been shown to be throughput-optimal in \cite{JW:convergence:08,liu-2009,Jiang-Shah-Shin-Walrand09}. This body of literature on completely distributed on-off scheduling has been extended to a framework that incorporates collisions in \cite{Jiang:contention:09,Ni-Tan-Srikant-09}. Further, this decentralized framework has been validated through experiments in \cite{Implementing-CSMA:09}.

However, independent set constraints can only model orthogonal channel access which, in general, is known to be sub-optimal \cite{Cover:2006fu} (Section $15.1$). For wireless networks, the interaction among nodes require a much more fine-grained characterization than independent set constraints. This can be fully captured in terms of the network's {\em rate region}, i.e., the set of link-rates that are simultaneously sustainable in the network. As long as the data-rates of links are within the {\em rate region}, simultaneous transmission is possible even by neighboring links in the network. Therefore, it is crucial to perform efficient distributed joint rate allocation (and not just distributed link scheduling) in wireless networks. Although distributed rate allocation is a very difficult problem in general, in this work, we show that this problem can be solved by taking advantage of physical-layer information.

In this work, we consider single-hop\footnote{For networks that do not employ cooperative schemes, the results in this paper are likely to generalize using multi-hop by combining ``back-pressure'' with the algorithmic framework of this paper.} wireless networks. We develop a simple, completely distributed algorithm for rate allocation in wireless networks that is throughput-optimal. In particular, given any rate region for a wireless network, we develop a decentralized (local queue-length based) algorithm that stabilizes all the queues for all arrival rates within the throughput region. Thus, we can utilize the entire physical-layer throughput region of the system with distributed rate allocation. To the best of our knowledge, this is the first paper to obtain such a result. This is a very exciting result as our decentralized algorithm achieves the same throughput region as optimal centralized cross-layer algorithms. The algorithm requires that each link can determine the global feasibility of increasing its data-rate from the current data-rate. In Section \ref{sec:feasibility}, we provide details on techniques to determine rate feasibility, and explain reasons for using this approach in practice.

The framework developed in this paper generalizes the distributed link scheduling framework. As discussed before, the current distributed link scheduling algorithms primarily deal with binary (on-off) decisions whereas our algorithm performs scheduling over multiple data-rates. Similar to these existing distributed link scheduling algorithms, our algorithm is mathematically modeled by a Markov process on the discrete set of data-rates. However, with multiple data-rates for each link, the appropriate choice of the large number of transition rates is very complicated. Thus, a key challenge is to design a Markov chain with fewer parameters that can be analyzed and appropriately chosen for throughput-optimality. We overcome this challenge by showing that transition rates with the following structure have this property. For link $i$, the transition rate to a data-rate $r_{i,j}$ from any other data-rate is $\exp(r_{i,j}v_i)$, where $v_i$ is a single parameter associated with link $i$ that is updated based on its queue-length. The transition takes place only if the new data-rate is feasible. As expected, this reduces to the existing algorithmic framework in the special case of binary (on-off) decisions.

For the general framework mentioned above, at an intuitive level, the techniques required for proving throughput-optimality remain similar to existing techniques. However, there are few additional technical issues that arise while analyzing the general framework. First, we need to account for more general constraints that arise from the set of possible rate allocation vectors. Next, the choice of update rules for $v_i(t)$ with time $t$ based on local queue-lengths that guarantee throughput-optimality does not follow directly. The mixing time of the rate allocation Markov chain plays an important role in choosing the update rules. For arbitrary throughput regions, any rate allocation algorithm that approach $\epsilon$-close (for arbitrarily small $\epsilon$) to the boundary possibly requires an increasing $1/\epsilon$ number of data-rates per link. This leads to a potential increase in the mixing time due to the increase in the size of the state-space. Thus, the analysis performed in this paper is more general and essential to establish throughput-optimality of the algorithms considered.

An important application of this algorithmic framework is for networks of white-space radios \cite{Supratim_09}, where multiple non-adjacent frequency bands are available for operation and multiple radios are available at the wireless nodes. A scheduler needs to allocate different radios to different bands in a distributed manner. This problem introduces multiple data-rates for every link even in the CSMA framework, and hence, existing distributed algorithms cannot be directly applied. We demonstrate that our framework provides a throughput-optimal distributed algorithm in this setting. 

Our main contributions are the following:
\begin{itemize}
\item We design a class of distributed cross-layer rate allocation algorithms for wireless networks that utilize local queue-length and physical-layer measuring. 
\item We show that there are algorithms in this class that are (\emph{a}) throughput-optimal, and (\emph{b}) completely decentralized.
\item We demonstrate that an adaptation of these algorithms are throughput-optimal for multiple-band multiple-radio distributed scheduling. 
\end{itemize}

\subsection{Notation}
Vectors are considered to be column vectors and denoted by bold letters. For a vector $\ba$ and matrix $B$, $\ba B := \ba^T B$, where $\ba^T$ is the transpose of $\ba$. For vectors, $\le$, $\ge$, $<$, $>$ and $=$ are defined component-wise. ${\bf 0}$ denotes all-zeros vector and ${\bf 1}$ denotes all-ones vector. Other basic notation used in the paper is given in Table \ref{tab:not}. Notation specific to proofs is introduced later as needed.
\begin{table}
\centering
\caption{Basic Notation}
\begin{tabular}{|c|l|} \hline
\label{tab:not}
$\bI(\cdot)$ & Indicator function\\ \hline
$\ba \cdot \bb$  & Dot product of vectors $\ba$ and $\bb$\\ \hline
$\|\ba\|_p$ & $L_p$-norm of vector $\ba$\\ \hline
$\|\ba\|_0$ & Number of non-zero elements of $\ba$\\ \hline
$|\cdot|$ &Absolute value for scalars,\\
&Cardinality for sets\\ \hline
$\bbE[\cdot]$ &Expectation operator\\ \hline
${\bbR_+}$ & Non-negative reals\\ \hline
${\bbZ_+}$ & Non-negative integers \\ \hline
${\bbZ_{++}}$ & Strictly positive integers \\ \hline
$\ei$ & Unit vector along $i$-th dimension, i.e.,\\
&$\ei \in \{0,1\}^n$ with $i$-th component equal to $1$\\
&and all other components equal to $0$\\ \hline
\end{tabular}
\end{table}

\subsection{Organization}
The next section describes the system model. Section \ref{sec:algorithm} explains the distributed rate allocation algorithm. Section \ref{sec:def_results} introduces relevant definitions and known results. Section \ref{sec:rate_stability} describes the rate allocation Markov chain and the optimization framework. Section \ref{sec:throughput_optimality} establishes the throughput-optimality of the algorithm. The algorithm for multiple-band multiple-radio scheduling is given in Section \ref{sec:application}. Further discussions and simulation results are given in Section \ref{sec:simulation}. We conclude with our remarks in Section \ref{sec:concl}. For readability, the proofs of the technical lemmas in Section \ref{sec:rate_stability} and Section \ref{sec:throughput_optimality} are moved to the Appendix.

\section{System Model}
\label{sec-system}

Consider a wireless network consisting of $m$ nodes, labeled $\cN :=\{1,2,\ldots,m\}$. In this network, we are interested in $n$ single-hop flows that correspond to $n$ wireless links labeled $\calL:=\{1, 2,\ldots,n\}$. Since we have a shared wireless medium, these links interact (or interfere) in a potentially complex way. For single-hop flows, this interaction among links can be captured through a $n$-dimensional \emph{rate region} for the network, which is formally defined next.
\begin{defn}[Rate Region]
The rate region of a network is defined as the set of instantaneous rate vectors $\bc \in \bbR_+^n $ at which queues (introduced later) of all $n$ links can be drained simultaneously.
\end{defn}
In this paper, we assume that the rate region is fixed\footnote{We consider fixed or slow-fading channels.} (i.e., not time-varying). We denote the rate region associated with the network by $\C \subseteq \bbR_+^n$. By definition, this rate region is compact. We assume that the rate region has the following simple property: if $\bc\in \C$, then $\hat{\bc}\in \C$ for all  $\hat{\bc}\le \bc$ and $\hat{\bc}\ge {{\bf 0}}$. The above property states that rates can be decreased component-wise. Such an assumption is fairly mild, and is satisfied by rate regions resulting from most physical-layer schemes. Next, we define the throughput region of the network.
\begin{defn}[Throughput Region]
The throughput region of a network, denoted by $\T$, is defined as the convex hull of the rate region $\C$ of the network. 
\end{defn}

We use a continuous-time model to describe system dynamics. Time is denoted by $t \in \bbR_+.$ Every (transmitter of) link $i \in \calL$ is associated with a queue $Q_i(t) \in \bbR_+$, which quantifies the information (packets) remaining at time $t$ waiting to be transmitted on link $i$. Let the cumulative arrival of information at the $i$-th link during the time interval $[0, t)$ be $A_i(t) \in \bbR_+$ with $A_i(0):=0$. \emph{Rate allocation} at time $t$ is defined as the rate vector in the rate region at which the system is being operated at time $t$. Let the rate allocation corresponding to the $i$-th link at time $t$ be $r_i(t)$. Then, for every link $i \in \calL$, the queue dynamics is given by
\be
\label{eq:queue}
Q_i(t) = Q_i(s) - \int_s^t r_i(z) \bI(Q_i(z)>0)dz + A_i(t) - A_i(s),
\ee
where $0\le s < t$. The vector of $n$ queues in the system is denoted by $\bQ(t):=[Q_i(t)]_{i=1}^{n}$. The queues are initially at $\bQ(0) \in \bbR_+^n$.

We consider arrival processes at the queues in the network with the following properties. 
\begin{itemize}
\item We assume every arrival process is such that increments over integral times are independent and identically distributed with $\text{Pr}(A_i(1)=0)>0.$
\item We assume that all these increments belong to a bounded support $[0, K]$, i.e., $A_i(k+1)-A_i(k) \in [0, K]$ for all $k \in \bbZ_+$.
\end{itemize}
Based on these properties, the (mean) \emph{arrival rate} corresponding to the $i$-th link is $\lambda_i:=\bbE[A_i(1)]$. We denote the vector of arrival rates by $\blam$. Without loss of generality\footnote{If $\lambda_i=0$, then this link can be removed from the system.}, we assume $\lambda_{\text{min}} := \min_i \lambda_{i} >0$. It follows from the strong law of large numbers that, with probability $1$,
\be
\label{eq:arr_strong_law}
\lim_{t\rightarrow \infty} \frac{A_i(t)}{t}=\lambda_i.
\ee

In summary, our system model incorporates general interference constraints through a arbitrary rate region and focuses on single-hop flows. We proceed to describe the rate allocation algorithm and the main results of this paper. 

\section{Rate Allocation Algorithm \& \\Main Results}
\label{sec:algorithm}

The goal of this paper is to design a completely decentralized algorithm for rate allocation that \emph{stabilizes} all the queues as long as the arrival rate vector is within the throughput region. By assumption, every link can determine rate feasibility, i.e., every link can determine whether increasing its data-rate from the current rate allocation results in a net feasible rate vector. More formally, every link $i \in \calL$ at time $t$, if required, can obtain the information $\bI(\br(t)+\alpha \ei \in \C), \text{ for any } \alpha \in \bbR.$ More details on determining rate feasibility are given in Section \ref{sec:simulation}.

The rate allocation vector at time $t$ is denoted by $\br(t):=[r_i(t)]_{i=1}^{n}$. For {\bf decentralized rate allocation}, we develop an algorithm that uses only local queue information for choosing $\br(t)$ over time $t$. Further, we perform rate allocation over a chosen limited (finite) set of rate vectors that are \emph{feasible}. We choose a finite set of rate levels corresponding to every link, and form vectors that are feasible. The details are as follows: 
\begin{enumerate}
\item For each link $i \in \calL$, a set of rate levels $\R_i=\{r_{i,j}\}_{j=0}^{k_i}$ are chosen from $[0, c_i]$ with $r_{i,0}=0$, $r_{i,k_i}=c_i$ and $r_{i,j-1} < r_{i,j}$.  Here, $c_i$ is the maximum possible transmission rate for the $i$-th link, i.e., $c_i := \arg\max_{\alpha \in \bbR_+} \alpha \ei \in \C$, and $k_i\in \bbZ_{++}$ is the number of levels other than zero. Since the rate region is compact, without loss of generality\footnote{If $c_i=0$, then this link can be removed from the system.}, we assume $0<\underline{K}\le c_{i}\le \bar{K}<\infty$. 
\item The set of rate allocation vectors, denoted by $\R$, is given by $\R =\{[r_1, r_2, \ldots, r_n] : r_i \in \R_i \text{ for all } i\in \calL, \text{ and } [r_1, r_2, \ldots, r_n] \in \C\}.$
\end{enumerate}

The convex hull of the set of rate allocation vectors $\R$ is denoted by $\R_c$. Define $\R_c^o =\{\br \in \bbR_+^n: \br < \bt \text{ for some } \bt \in \R_c\}, $ the set of \emph{strictly} feasible rates.  For rate regions that are polytopes, the partitions $\R_i$ can be chosen such that $\R_c = \T$. For any compact rate region, it is fairly straightforward to choose partitions $\R_i$ with $k_i \le \left\lceil {2c_i}/{\epsilon}\right\rceil \le \left\lceil {2\bar{K}}/{\epsilon}\right\rceil$ such that $\bc \in \R_c$ if $\bc+\frac{\epsilon}{2} {\bf 1} \in \T$. The trivial partition with $\epsilon/2$ as step size in all dimensions satisfy the above property. Thus, for any given $\epsilon>0$, we can obtain a set of rate allocation vectors $\R$ such that 
\be
\label{eq:card_bound}
|\R|\le \left\lceil {2\bar{K}}/{\epsilon}\right\rceil^n
\ee
and $\bc \in \R_c$ if $\bc+\frac{\epsilon}{2} {\bf 1} \in \T$.

Before describing the algorithm, we define two notions of throughput performance of a rate allocation algorithm.

\begin{defn}[Rate stable] We say that a rate allocation algorithm is rate-stable if, for any $\blam \in \R_c^o$, the departure rate corresponding to every queue is equal to its arrival rate, i.e., for all $i \in \calL$, with probability $1$,
\[
\lim_{t \rightarrow \infty}\frac{1}{t}\int_0^t r_i(z) \bI(Q_i(z)>0)dz = \lambda_i.
\] From (\ref{eq:queue}),(\ref{eq:arr_strong_law}), this is same as, for all $i \in \calL$, with probability $1$,
$$
\lim_{t \rightarrow \infty}Q_i(t)/t = 0.
$$
\end{defn}
\begin{defn}[Throughput optimal] We say that a rate allocation algorithm is throughput-optimal if, for any given $\epsilon > 0$, the algorithm makes the underlying network Markov chain \emph{positive Harris recurrent} (defined in Section \ref{sec:def_results}) for all $\blam$ such that $\blam + \epsilon {\bf 1}\in \T$. By definition, the algorithm can depend on the value of $\epsilon.$
\end{defn}

Next, we describe {\bf a class of algorithms} to determine $\br(t)$ as a function of time based on a continuous-time Markov chain. Recall that $\R_i=\{r_{i,j}\}_{j=0}^{k_i}$ is the set of possible rates/states for allocation associated with the $i$-th link. In these algorithms, the $i$-th link uses $k_i$ independent exponential clocks with rates/parameters\footnote{These should not to be confused with the rates for allocation.} $\{U_{i,j}\}_{j=0}^{k_i}$ (or equivalently exponential clocks with mean times $\{{1}/{U_{i,j}}\}_{j=0}^{k_i}$). The clock with (time varying) parameter $U_{i,j}$ is associated with the state $r_{i,j}$. Based on these clocks, the $i$-th link obtains $r_i(t)$ as follows: 
\begin{enumerate}
\item If the clock associated with a state (say $j=m$) ticks and further if transitioning to that state $r_{i,m}$ is feasible, then $r_i(t)$ is changed to $r_{i,m}$; 
\item Otherwise, $r_i(t)$ remains the same. 
\end{enumerate}
The above procedure continues, i.e, all the clocks run continuously. Define $u_{i,j}:=\log U_{i,j}, \forall i \in \calL, j \in \{0, 1, \ldots, k_i\}$. It turns out that the appropriate structure to introduce is as follows:
$$u_{i,j}= r_{i,j} v_i, \forall i \in \calL, j \in \{0, 1, \ldots, k_i\},$$
where $v_i \in \bbR, \forall i \in \calL.$ We denote the vector consisting of these new set of parameters by $\bv:=[v_i]_{i=1}^n$.

\begin{figure}[!t]
\centering
\scalebox{1}{\input{mac_channel.pstex_t}}
\caption{Gaussian Multiple Access Channel}
\label{fig:MAC_channel}
\end{figure}
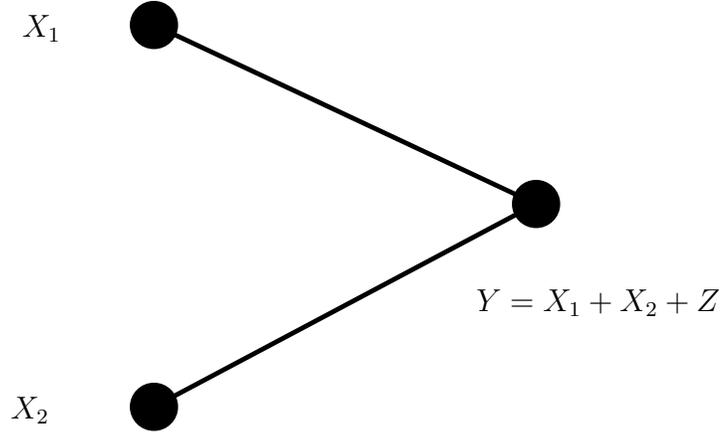

\begin{figure}
\centering
\epsfig{file=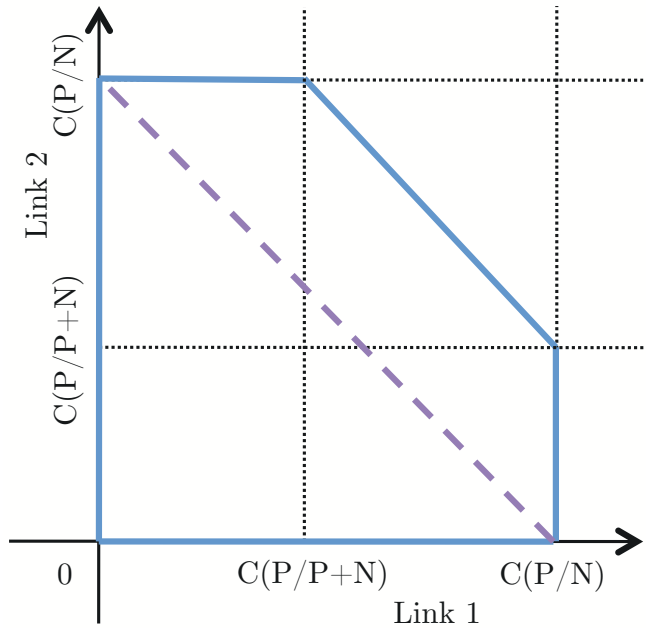,width=3.5in}
\caption{Information-theoretic Capacity Region}
\label{fig:example}
\end{figure}

\begin{figure}
\centering
\epsfig{file=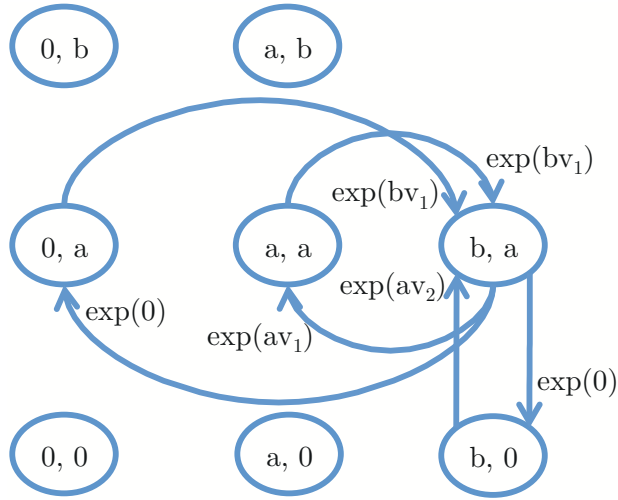,width=3.5in}
\caption{Rate Allocation Markov Chain (transitions to/from $(b,a)$ state alone shown)}
\label{fig:markov_chain}
\end{figure}

\begin{exm}
Consider a Gaussian multiple access channel with two links as shown in Figure \ref{fig:MAC_channel} with average power constraint $P$ at the transmitters and noise variance $N$ at the receiver. The capacity region of this channel is shown in Figure \ref{fig:example} where $C(x)=0.5\log_2(1+x)$. In this case, orthogonal access schemes limit the throughput region to the triangle (strictly within the pentagon) shown using dash-line. In this example, if we allow for capacity-achieving physical-layer schemes, the rate region (and hence the throughput region) is identical to the pentagon shown in Figure \ref{fig:example}. The natural choice for the set of rate levels at link-1 is $\R_1=\{0, a, b\}$ where $a=C(P/P+N)$ and $b=C(P/N)$. Similarly, $\R_2=\{0, a, b\}$. This leads to the set of rate allocation vectors $\R=\{[0, 0], [0, a], [0, b], [a, 0], [a, a], [a, b], [b, 0], [b, a]\}.$ It is clear that the convex combination of this set is the throughput region itself.  For this example, the state-space of the Markov chain and transitions to and from state $(b,a)$ are shown in Figure \ref{fig:markov_chain}.
\end{exm}

A distributed algorithm needs to choose the parameters $\bv$ in a decentralized manner. For providing the intuition behind the algorithm, we perform this in two steps. In the first step, we develop the non-adaptive version of the algorithm that has the knowledge of $\blam$. This algorithm is called non-adaptive as the algorithm requires the explicit knowledge of $\blam$. The rate allocation at time $t=0$ is set to be $\br(0)={\bf 0}$. This algorithm uses $\bv^*$ at all times which is a function of $\blam$, and is given by
\[
\bv^* = \arg\max_{\bv \in \bbR^n} \blam \cdot \bv - \log \left(\sum_{\br \in \R} \exp\left(\br \cdot \bv\right) \right).
\]
We show in Section \ref{sec:rate_stability} that, given $\blam \in \R_c^o$, the above optimization problem has a unique solution that is finite, and therefore has a valid $\bv^*$. An important result regarding this non-adaptive algorithm is the following theorem. 

\begin{theorem}
\label{thm:rate_stability}
The above non-adaptive algorithm is rate-stable for any given $\blam \in \R_c^o$.
\end{theorem}
\begin{IEEEproof}[Proof Outline] For any $\blam \in \R_c^o$, there is at least one distribution on $\R$ that has expectation as $\blam$. For the Markov chain specified by any $\bv\in \bbR^n$, there is a stationary distribution on the state-space $\R$. The value $\bv^*$ is chosen such that it minimizes the Kullback-Leibler divergence of the induced stationary distribution from the distribution corresponding to $\blam$. For the Markov chain specified by $\bv^*$, the expected value of the stationary distribution turns out to be $\blam$. This leads to rate-stable performance of the algorithm. The proof details are given in Section \ref{sec:rate_stability}. 
\end{IEEEproof}

In the second step, we develop the {\bf adaptive algorithm}, where $\bv$ is obtained as a function of time $t$ denoted by $\bv(t)\footnote{This implies that the exponential clocks used have time varying rates. These are well-defined non-homogeneous Poisson processes.}.$ This algorithm is called adaptive as the algorithm does not require the knowledge of $\blam$. The values of $\bv(t)$ are updated during fixed (not random variables) time instances $\tau_l$ for $l \in \bbZ_{++}$. We set $\tau_{0}=0$ and $\bv(0)={\bf 0}$. During interval $t\in[\tau_l, \tau_{l+1})$ the algorithm uses $\bv(t)=\bv(\tau_l)$. The length of the intervals are $T_l=\tau_{l+1}-\tau_l$. During interval $[\tau_l, \tau_{l+1})$, let the \emph{empirical arrival rate} be 
\be
\label{eq:lamhat_defn}
\hat{\lambda}_i(l)=\frac{A_i(\tau_{l+1}) - A_i(\tau_l)}{T_l}
\ee
and the \emph{empirical offered service rate} be
\be
\label{eq:shat_defn}
\hat{s}_i(l)=\frac{1}{T_l}\int_{\tau_l}^{\tau_{l+1}} r_i(z) dz.
\ee
The update equation corresponding to the algorithm for the $i$-th link is given by
\be
\label{eq:grad_step}
v_i(\tau_{l+1}) = \left[v_i(\tau_l)+\alpha_l\left(\hat{\lambda}_i(l) + \frac{\epsilon}{4} -\hat{s}_i(l)\right)\right]_{D}
\ee
where $[\theta]_D = \min(\theta,D)\bI(\theta\ge 0)+\max(\theta,-D)\bI(\theta < 0)$, i.e., $[\theta]_D$ is the projection of $\theta$ to the closest point in $[-D,D]$, and $\alpha_l$ are the step sizes. Thus, the algorithm parameters are interval lengths $T_l$, step sizes $\alpha_l$ and $D$. 
\begin{remark}
Clearly, both empirical arrival rate and empirical offered service rate used in the above algorithm can be computed by the $i$-th link without any external information. In fact, the difference is simply the difference of its queue-length over the previous interval appropriately scaled by the inverse of the length of the previous interval.
\end{remark}


The following theorem provides $\epsilon$-optimal performance guarantee for the adaptive algorithm.

\begin{theorem}
\label{thm:throughput_optimal}
Consider any given $\epsilon>0$, $\epsilon \le 4 \lambda_{\text{\emph{min}}}$. Then, there exists some choice of algorithm parameters $T_l=T(n,\epsilon)$, $\alpha_l=\alpha(n,\epsilon)$ and $D=D(n,\epsilon)$ such that the appropriate network Markov chain under the adaptive algorithm is positive Harris recurrent if $\blam + \epsilon {\bf 1} \in \T$, i.e., the algorithm is throughput-optimal.
\end{theorem}

\begin{IEEEproof}[Proof Outline] The update in (\ref{eq:grad_step}) can be intuitively thought of as a gradient decent technique to solve an optimization problem that will lead to $\bv^*$ whose induced stationary distribution on $\R$ has expected value \emph{strictly} greater than $\blam$. However, the arrival rate and offered service rate are replaced with their empirical values for decentralized operation. We consider the two time scales involved in the algorithm - update interval $T$ and $N$ update intervals. The main steps involved in establishing the throughput-optimality are the following. First, we show that, sufficiently long $T$ can be chosen such that the empirical values used in the algorithm are arbitrarily close to the true values. Using this, we next show that the average offered empirical service rate over $N$ update intervals is \emph{strictly} higher than the arrival rate. Finally, we show that this results in a \emph{drift} that is sufficient to guarantee positive Harris recurrence. The proof details are given in Section \ref{sec:throughput_optimality}. 
\end{IEEEproof}

\section{Definitions \& Known Results}
\label{sec:def_results}

We provide definitions and known results that are key in establishing the main results of this paper. We begin with definitions on two measures of difference between two probability distributions.
\begin{defn}[Kullback-Leibler (KL) divergence] Consider two probability mass functions $\pi$ and $\mu$ on a finite set $\X$. Then, the KL divergence from $\pi$ to $\mu$ is defined as $D(\mu\|\pi)=\sum_{\bx\in\X}\mu(\bx) \log \frac{\mu(\bx)}{\pi(\bx)}.$
\end{defn}
\begin{defn}[Total Variation]Consider two probability mass functions $\pi$ and $\mu$ on a finite set $\X$. Then, the total variation distance between $\pi$ and $\mu$ is defined as $\|\mu-\pi\|_{TV}=\frac{1}{2}\sum_{\bx\in\X}|\mu(\bx)-\pi(\bx)|.$
\end{defn} 


Next, we provide two known results that are used later. Result \ref{res:mix_time} follows directly from \cite{Bremaud-99}(Theorem $3.2$), and Result \ref{res:conductance} is in \cite{Bremaud-99}(Theorem $4.3$).

\begin{result}[Mixing Time]
\label{res:mix_time}
Consider any finite state-space, aperiodic, irreducible, discrete-time Markov chain with transition probability matrix $P$ and the stationary distribution $\balpha$. Let  $\alpha_{\text{\emph{min}}}$ be the minimum value in $\balpha$ and the second largest eigenvalue modulus (SLEM) be $\sigma_{\text{\emph{max}}}$. Then, for any $\rho > 0$, starting from any initial distribution (at time 0), the distribution at time $\tau \in \bbZ_{++}$ associated with the Markov chain, denoted by $\bbeta(\tau)$, is such that $\|\bbeta(\tau)-\balpha\|_{TV}\le \rho$ if 
\be
\label{eq:result_JSSW}
\tau \ge \frac{\frac{1}{2}\log(1/\alpha_{\text{\emph{min}}})+\log(1/\rho)}{\log(1/\sigma_{\text{\emph{max}}})}.
\ee
\end{result}

\begin{result}[Conductance Bounds]
\label{res:conductance}
Consider the setting as above. The ergodic flow out of $\calS \subseteq \X$ is defined as $F(\calS):=\sum_{\bx\in \calS, \hat{\bx} \in \calS^c} \balpha(\bx)P(\bx,\hat{\bx})$ and the conductance is defined as
\be
\label{eq:conductance_def}
\Phi = \min\left\{\frac{F(\calS)}{\sum_{\bx\in \calS}\balpha(\bx)}:\phi\subset \calS\subset \X, \sum_{\bx\in \calS}\balpha(\bx) \le \frac{1}{2}\right\}.
\ee Then, the SLEM $\sigma_{\text{\emph{max}}}$ is bounded by conductance as follows:
\be
\label{eq:cheegers}
1-2\Phi \le \sigma_{\text{\emph{max}}} \le 1 - \Phi^2/2.
\ee  
\end{result}

Lastly, we provide the definition of positive Harris recurrence. For details on properties associated with positive Harris recurrence, see \cite{Meyn-Tweedie-93,Dai_95}. 
\begin{defn}[Positive Harris recurrence] Con-sider a discrete-time time-homogeneous Markov chain on a complete, separable metric space $X$. Let $\cal{B}_X$ denote the Borel $\sigma$-algebra on $X$, and $X_\tau$ denote the state of the Markov chain at time $\tau \in \bbZ_+$. Define stopping time $T_A:=\inf\{\tau \ge 0: X_\tau \in A\}$ for any $A \in \cal{B}_X$. The set $A$ is called Harris recurrent if $\text{\emph{Pr}}(T_A<\infty|X(0)=x)=1$ for any $x\in X$. A Markov chain is called Harris recurrent if there exits a $\sigma$-finite measure $\mu$ on $(X,\cal{B}_X)$ such that if $\mu(A)>0$ for some $A\in \cal{B}_X$, then $A$ is Harris recurrent. It is known that if $X$ is Harris recurrent an essentially unique invariant measure exists. If the invariant measure is finite, then it may be normalized to a probability measure. In this case, $X$ is called positive Harris recurrent.
\end{defn}
 
\section{Rate allocation Markov chain \& Rate Stability}
\label{sec:rate_stability}

{\bf Rate allocation Markov chain:} The main challenge is to design a Markov chain with fewer parameters that can be analyzed and appropriately chosen for throughput-optimality. First, we identify a class of Markov chains that are relatively easy to analyze. Consider the class of algorithms introduced in Section \ref{sec:algorithm}. The core of this class of algorithms is a continuous-time Markov chain with state-space $\R$, which is the (finite) set of rate allocation vectors. Define
\be
\label{eq:fun_f}
f(\hat{\br},\br):=\exp\left(\sum_{i=1}^{n}\sum_{j=0}^{k_i}u_{i,j}\bI(r_i ={r}_{i,j})\bI(r_i \ne \hat{r}_i)\right),
\ee
where $\hat{\br}=[\hat{r}_1, \hat{r}_2, \ldots, \hat{r}_n] \in \R$, $\br=[r_1, r_2, \ldots, r_n]  \in \R$ and $u_{i,j}$ are the parameters introduced in Section \ref{sec:algorithm}. Now, the transition rate from state $\hat{\br}\in \R$ to state $\br \in \R$ can be expressed as 
\[
q(\hat{\br},\br)=\left\{
\begin{array}{ll}
f(\hat{\br},\br), &\text{if } \|\hat{\br}-\br\|_0 = 1,\\
0, &\text{if } \|\hat{\br}-\br\|_0 > 1.\\
\end{array}
\right.
\]  
And, the diagonal elements of the rate matrix are given by $q(\hat{\br},\hat{\br}) = -\sum_{\br \in \R, \br \ne \hat{\br}} q(\hat{\br},\br)$ for all $\hat{\br} \in \R$. This follow directly from the description of the algorithm. This class of algorithms are carefully designed such that it is tractable for analysis. In particular, the following lemma shows that this Markov chain is reversible and the stationary distribution has exponential form.

\begin{lemma}
\label{lem:finite_state_MC}
The rate allocation Markov chain $(\R,q)$ is reversible and has the stationary distribution
\be \label{eq:st_dist}
\pi(\br)= \frac{\exp\left(\sum_{i=1}^{n}\sum_{j=0}^{k_i}u_{i,j}\bI(r_i ={r}_{i,j})\right)}{\sum_{\tilde{\br} \in \R}\exp\left(\sum_{i=1}^{n}\sum_{j=0}^{k_i}u_{i,j}\bI(\tilde{r}_i ={r}_{i,j})\right)}.
\ee
Furthermore, this Markov chain converges to this stationary distribution starting from any initial distribution.
\end{lemma}

\begin{IEEEproof}
The proof follows from detailed balance equations $\pi({\br})q(\br,\hat{\br})=\pi({\hat{\br}})q(\hat{\br},\br)$ for all $\br, \hat{\br} \in \R$ and known results on convergence to stationary distribution for irreducible finite state-space continuous-time Markov chains \cite{B:A03}.
\end{IEEEproof}

The \emph{offered service rate} vector under the stationary distribution is $\bs := \sum_{\br \in \R} \pi(\br) \br$. In general, for $\blam \in \R_c^o$, we expect to find values for parameters $u_{i,j}$ as a function of $\blam$ and $\R$ such that $\bs=\blam$. Due the exponential form in (\ref{eq:st_dist}), it turns out that the right structure to introduce is
\be \label{eq:structure}
u_{i,j}= r_{i,j} v_i, \forall i \in \calL, j \in \{0, 1, \ldots, k_i\},
\ee
where $v_i \in \bbR, \forall i \in \calL$, and obtain suitable values for $\bv=[v_i]_{i=1}^n$ as a function of $\blam$ and $\R$ such that $\bs=\blam$. To emphasize the dependency on $\bv$, from now onwards, we denote the stationary distribution by $\pi_\bv(\br)$ and the offered service rate vector by 
\be
\label{eq:service_vector}
\bs_\bv = \sum_{\br \in \R} \pi_\bv(\br) \br.
\ee Substituting (\ref{eq:structure}), we can simplify (\ref{eq:st_dist}) to obtain
\be \label{eq:st_dist_simple}
\pi_\bv (\br)= \frac{\exp(\br \cdot \bv)}{\sum_{\tilde{\br} \in \R}\exp(\tilde{\br} \cdot \bv)}.
\ee

{\bf Optimization framework:} We utilize the optimization framework in \cite{Jiang-Allerton} to show that values for $\bv$ exist such that $\bs_\bv=\blam$. In particular, we show that the unique solution to an optimization problem given by $\bv^*$ has the property $\bs_{\bv^*}=\blam$. Next, we describe the intuitive steps to arrive at the optimization problem. If $\blam \in \R_c^o$, then $\blam$ can be expressed as a convex combination of ${\br \in \R}$, i.e., there exists a valid probability distribution $\mu(\br)$ such that $\blam = \sum_{\br \in \R} \mu(\br) \br$. For a given distribution $\mu(\br)$, we are interested in choosing $\bv$ such that $\pi_\bv(\br)$ is \emph{close} to $\mu(\br)$. We consider the KL divergence of $\pi_\bv(\br)$ from $\mu(\br)$ given by $D\left(\mu(\br)\|\pi_\bv(\br)\right)$. Minimizing $D\left(\mu(\br)\|\pi_\bv(\br)\right)$ over the parameter $\bv$ is equivalent in terms of the optimal solution(s) to maximizing 
$F(\mu(\br),\pi_\bv(\br)):=-D\left(\mu(\br)\|\pi_\bv(\br)\right)-H(\mu(\br))$ over the parameter $\bv$ as $H(\mu(\br))$ is a constant. Simplifying $F(\mu(\br),\pi_\bv(\br))$ leads the optimization problem as follows:
\be
F(\mu(\br),\pi_\bv(\br))&=&\sum_{\br \in \R} \mu(\br) \log \pi_\bv(\br) \n \\
&\overset{(a)}{=}& \sum_{\br \in \R} \mu(\br) \br \cdot \bv - \log\left(\sum_{\br \in \R}\exp(\br \cdot \bv)\right) \n \\
&\overset{(b)}{=}& \blam \cdot \bv - \log\left(\sum_{\br \in \R}\exp(\br \cdot \bv)\right).\n
\ee
Here, $(a)$ follows from (\ref{eq:st_dist_simple}) and $(b)$ follows from the assumption $\blam = \sum_{\br \in \R} \mu(\br) \br$. Now onwards, we denote the objective function by $F(\bv,\blam)$. To summarize, the optimization problem of interest is, given  $\blam \in \R_c^o$,
\be \label{eq:opt_problem}
\text{maximize} &F(\bv,\blam)=\blam \cdot \bv - \log\left(\sum_{\br \in \R}\exp(\br \cdot \bv)\right)\\
\text{subject to} &\bv \in \bbR^n. \n
\ee

The following lemma regarding the optimization problem in (\ref{eq:opt_problem}) is a key ingredient to the main results.
\begin{lemma}
\label{lem:rate_stable}
Let $\blam \in \R_c^o$. The optimization problem in (\ref{eq:opt_problem}) has a unique solution $\bv^*(\blam)$, which is finite. In addition, the {offered service rate} vector under $\bv^*$ is equal to the arrival rate vector, i.e., $\bs_{\bv^*}= \blam.$
\end{lemma}
\begin{IEEEproof}
See Appendix.
\end{IEEEproof}
The important observations are that the objective function is concave in $\bv$ and the gradient with respect to $\bv$ is $\blam - \bs_{\bv}$. With offered service rate equal to arrival rate, the next step is to show that the queues drain at rate equal to $\blam$.

\subsection{Proof of Theorem \ref{thm:rate_stability}}

{\bf Rate stability of the non-adaptive algorithm:} We establish the rate stability of the non-adaptive algorithm with the result given in Lemma \ref{lem:rate_stable} as follows.

Consider time instances $\nu_l$ for $l \in \bbZ_+$ with $\nu_{0}=0$, and interval length $\Gamma_l:=\nu_{l+1}-\nu_l=l+1$. The queue at the $i$-th link can be upper bounded as follows. The offered service during the time interval is $[\nu_{k}, \nu_{k+1})$ is used to serve the arrivals during the time interval $[\nu_{k-1}, \nu_{k})$ \emph{alone}. Consider a time $t$, and choose $l$ such that $t\in [\nu_l, \nu_{l+1})$. Using (\ref{eq:queue}) and the above upper bounding technique, we obtain
\be
Q_i(t) &=& A_i(t) - \int_0^t r_i(z) \bI(Q_i(z)>0)dz \n \\
&\le& \sum_{k=0}^{l-2}\left[A_i(\nu_{k+1}) - A_i(\nu_{k}) - \int_{\nu_{k+1}}^{\nu_{k+2}} r_i(z) dz\right]_+\n \\
\label{eq:queue_upperbound}
&&+A_i(t) - A_i(\nu_{l-1}),
\ee
where $[\theta]_+ = \max(0,\theta).$

For each interval $[\nu_{k}, \nu_{k+1})$, define the following two random variables:
\[
\alpha_i(k):=\frac{A_i(\nu_{k+1}) - A_i(\nu_{k})}{\Gamma_k}, \text{ and}
\]
\[
\beta_i(k):=\frac{1}{\Gamma_{k}}\int_{\nu_{k}}^{\nu_{k+1}} r_i(z) dz.
\]
It follows from the strong law of large numbers that, with probability $1$, $\lim_{k\rightarrow \infty} \alpha_i(k) = \lambda_i$. From Lemma \ref{lem:rate_stable} and ergodic theorem for Markov chains, it follows that, with probability $1$, $\lim_{k\rightarrow \infty} \beta_i(k+1) = \lambda_i.$ Since the arrival process $A_i(t)$ is non-decreasing and the increments are bounded by $K$, we have
\be
A_i(t) - A_i(\nu_{l-1}) &\le& A_i(\nu_{l+1}) - A_i(\nu_{l-1}) \n \\
&\le& K(\nu_{l+1}- \nu_{l-1}) \n \\
\label{eq:tail_bound}
&=& K(\Gamma_{l-1}+ \Gamma_l).
\ee
Rewriting (\ref{eq:queue_upperbound}) with above defined random variables and applying (\ref{eq:tail_bound}) along with $\nu_l\le t$ and $\Gamma_k\le \Gamma_{k+1}$, we obtain
\be
\frac{Q_i(t)}{t} &\le& \frac{1}{\nu_l}\sum_{k=0}^{l-2}\Gamma_k\left[\alpha_i(k)-\beta_i(k+1)\right]_+ \n\\
\label{eq:queue_UB_fin}
&&+ \frac{K(\Gamma_{l-1}+ \Gamma_l)}{\nu_l}.
\ee
In (\ref{eq:queue_UB_fin}), the second term on the right hand side (RHS) goes to zero as $\Gamma_l/\nu_l \rightarrow 0$ as $l \rightarrow \infty$. The first term on the RHS of (\ref{eq:queue_UB_fin}) goes to zero with probability $1$ as $\alpha_i(k)-\beta_i(k+1) \rightarrow 0$, $\nu_l\ge \sum_{k=0}^{l-2}\Gamma_k$ and $\nu_l\rightarrow \infty.$ Thus, for any given $\blam \in \R_c^o$, with probability $1$,
$$\lim_{t\rightarrow \infty} \frac{Q_i(t)}{t}=0, \forall i \in \calL,$$ which completes the proof.
\hfill $\IEEEQED$

This result is important due to the following two reasons. 
\begin{enumerate}
\item The result shows that this algorithm has good performance, and an algorithm that approaches the operating point of this algorithm has the potential to perform ``well.'' Essentially, this aspect is utilized to obtain the adaptive algorithm. 
\item The non-adaptive algorithm does not require the knowledge of the number of nodes or $\epsilon$, as required by the adaptive algorithm. This suggests the existence of similar gradient-like algorithms that perform ``well'' with different algorithm parameters that may not depend on the number of nodes or $\epsilon$. We do not address this question in the paper, but the non-adaptive algorithm will serve as the starting point to address such issues.
\end{enumerate}

\section{Throughput Optimality of Algorithm}
\label{sec:throughput_optimality}

In this section, we establish the throughput-optimality of the adaptive algorithm for a particular choice of parameters. The algorithm parameters used in this section are dependent on the number of links $n$ and $\epsilon$. It is evident from the theorem that $\epsilon$ determines how close the algorithm is to optimal performance. Define $$C(n):=3^5(2\bar{K}+K)^2\left(\frac{\bar{K}^2n^2}{2}+n\right).$$ We set all the step sizes (irrespective of interval) to 
\be
\label{eq:alpha}
\alpha_l=\alpha(n,\epsilon) :={\epsilon^2}/{C(n)},
\ee and $D$ used in the projection to
\be
\label{eq:D}
D=D(n,\epsilon):= \frac{16\bar{K}}{\underline{K}}\frac{n}{\epsilon}\log\left\lceil\frac{2\bar{K}}{\epsilon}\right\rceil + \bar{K}.
\ee
All the interval lengths (irrespective of interval) are set to
\be
\label{eq:T}
T_l=T(n,\epsilon):=\exp\left(\hat{K}\left(\frac{n^2}{\epsilon}\log\frac{n}{\epsilon}\right)\right)
\ee
for some large enough constant $\hat{K}>0$.

\begin{remark}
The large value of $T(n,\epsilon)$ in (\ref{eq:T}) is due to the poor bound on the conductance of the rate allocation Markov chain. The parameters given by (\ref{eq:alpha}), (\ref{eq:D}) and (\ref{eq:T}) are one possible choice of the parameters. We would like to emphasize that this choice is primarily for the purpose of the proofs. The choice of right parameters (and even the update functions) in practice are subject to further study especially based on the network configuration and delay requirements. Some comments on this are given in Section \ref{sec:simulation}.
\end{remark}

We start with the optimization framework developed in the previous section. For the adaptive algorithm, the relevant optimization problem is as follows: given $\blam$ such that $\blam + \frac{\epsilon}{2}{\bf 1}\in \R_c$,
\be \label{eq:opt_problem_2}
&\text{maximize} &F_{\epsilon}(\bv):=F\left(\bv,\blam+\frac{\epsilon}{4}{\bf 1}\right)\\
&\text{subject to} &\bv \in \bbR^n. \n
\ee
The following result is an extension of Lemma $\ref{lem:rate_stable}$. 
\begin{lemma}
\label{lem:opt_prob_2}
Consider any given $\epsilon > 0$ and $\blam$. Then, the optimization problem in (\ref{eq:opt_problem_2}) is strictly concave in $\bv$ with gradient $\nabla F_{\epsilon}(\bv)=\blam+\frac{\epsilon}{4}{\bf 1} - \bs_{\bv}$ and Hessian 
\[H(F(\bv))= - \left(\bbE_{\pi_\bv}\left[\br \br^T\right]-\bbE_{\pi_\bv}\left[\br\right]\bbE_{\pi_\bv}\left[\br^T\right]\right).\] Further, let $\blam + \frac{\epsilon}{2}{\bf 1}\in \R_c$. Then, it has a unique solution $\bv^*$, which is finite, such that the {offered service rate} vector under $\bv^*$ is equal to $\blam+\frac{\epsilon}{4}{\bf 1}$, i.e., $\bs_{\bv^*}=\blam+\frac{\epsilon}{4}{\bf 1}.$
In addition, if $\epsilon \le 4 \lambda_{\text{\emph{min}}}$, then the optimal value $\bv^*$ is such that
\be
\label{eq:vinfbound}
\|\bv^*\|_{\infty} \le \frac{16\bar{K}}{\underline{K}}\frac{n}{\epsilon}\log\left\lceil\frac{2\bar{K}}{\epsilon}\right\rceil.
\ee
\end{lemma}
\begin{IEEEproof}
See Appendix.
\end{IEEEproof}

The update step in (\ref{eq:grad_step}), which is central to the adaptive algorithm, can be intuitively thought of as a gradient decent technique to solve the above optimization problem. Technically, it is different as the arrival rate and offered service rate are replaced with their empirical values for decentralized operation. The algorithm parameters can be chosen in order to account for this. This forms the central theme of this section. 

\subsection{Within update interval}
Consider a time interval $[\tau_l, \tau_{l+1})$. During this interval the algorithm uses parameters $v_i(\tau_{l})$. For simplicity, in this subsection, we denote $v_i(\tau_{l})$ by $v_i$ and the vector by $\bv$ and $\bbE[\cdot|\bv]$  by $\bbE[\cdot]$. For the rate allocation Markov chain (MC) introduced in Section \ref{sec:rate_stability}, we obtain an upper bound on the convergence time or the mixing time.

To obtain this bound, we perform \emph{uniformization} of the CTMC (continuous-time MC) and use results given in Section \ref{sec:def_results} on the mixing time of DTMC (discrete-time MC). The uniformization constant used is $A:=n\exp(\bar{K}\|\bv\|_{\infty})$. The resulting DTMC has the same state-space $\R$ with transition probability matrix $P$. The transition probability from state $\hat{\br}\in \R$ to state $\br\in \R,\br \ne \hat{\br}$ is $P(\hat{\br},\br)=q(\hat{\br},\br)/A$, and from state $\hat{\br}\in \R$ to itself is $P(\hat{\br},\hat{\br})=1+q(\hat{\br},\hat{\br})/A$. With our choice of parameters $u_{i,j}$ given by (\ref{eq:structure}), we can simplify (\ref{eq:fun_f}) to \be
\label{eq:simpl_f}f(\hat{\br},\br)=\exp\left(\sum_{i=1}^{n}r_{i}v_{i}\bI(r_i \ne \hat{r}_i)\right).
\ee For all $\hat{\br},\br\in \R,\br \ne \hat{\br}$, clearly $q(\hat{\br},\br)\le \exp(\bar{K}\|\bv\|_{\infty})$. Since at most $n$ elements in every row of the transition rate matrix of the CTMC is positive $|q(\hat{\br},\hat{\br})|\le A$ for all $\hat{\br}\in \R$. Therefore, $P$ is a valid probability transition matrix. 

The DTMC has the same stationary distribution as the CTMC. In addition, the CTMC and the DTMC have one-to-one correspondence through an underlying independent Poisson process with rate $A.$ In this subsection, time $t$ denotes the time within the update interval, i.e., $t=0$ denotes global time $\tau_l$. Let $\bmu(t)$ be the distribution over $\R$ given by the CTMC at time $t$, and $\zeta$ be a Poisson random variable with parameter $At$. Then, we have
\be
\bmu(t)&=&\sum_{m \in \bbZ_+}\text{Pr}(\zeta=m)\bmu(0)P^m\n\\
\label{eq:mu_dist_t}
&=&\bmu(0)\exp(At(P-I)),
\ee
where $I$ is the identity matrix. Next, we provide the upper bound on the mixing time of the CTMC.

\begin{lemma}
\label{lem:mix_time_bound}
Consider any $\rho_1>0$. Then, there exists a constant $K_1>0$, such that, if 
\be
\label{eq:mix_time_bound}
t \ge  \exp\left(K_1\left(n\|\bv\|_{\infty}+n\log\frac{1}{\epsilon}\right)\right)\log \frac{1}{\rho_1},
\ee
then the total variation between the probability distribution $\bmu(t)$ at time $t$ given by (\ref{eq:mu_dist_t}) and the stationary distribution $\bpi_\bv$ given by (\ref{eq:st_dist_simple}) is smaller than $\rho_1$, i.e., $\|\bmu(t)-\bpi_\bv\|_{TV}\le \rho_1.$
\end{lemma}
\begin{IEEEproof}
See Appendix.
\end{IEEEproof}

Lemma \ref{lem:mix_time_bound} is used to show that the error associated with using empirical values for arrival rate and offered service rate in the update rule (\ref{eq:grad_step}) can be made arbitrarily small by choosing large enough $T$. This is formally stated in the next lemma.

\begin{lemma}
\label{lem:error_bound}
Consider $\rho_2>0$. Then, there exists a constant $K_2>0$, such that, if the updating period 
\[
T \ge \exp\left(K_2\left(n\|\bv\|_{\infty}+n\log\frac{1}{\epsilon}\right)\right)\frac{1}{\rho_2},
\] then for any time interval $[\tau_l, \tau_{l+1})$
\be
\label{eq:error_bound_T}
\bbE\left[\left\|\hat{\blam}(l) -{\blam}\right\|_1\right]+\bbE\left[\left\|\hat{\bs}(l) -{\bs_{\bv}}\right\|_1\right] \le \rho_2.
\ee
\end{lemma}
\begin{IEEEproof}
See Appendix.
\end{IEEEproof}

Thus, the important result is that due to the mixing of the rate allocation Markov chain, the empirical offered service rate is \emph{close} to the offered service rate. The next step is to address whether the offered service rates over multiple update intervals is \emph{higher} than the arrival rates.

\subsection{Over multiple update intervals}

We consider multiple update intervals, and establish that the average empirical offered service rate is \emph{strictly} higher than the arrival rate. This result follows from the observation that, if the error in approximating the true values by empirical values are sufficiently small, then the expected value of the gradient of $F_{\epsilon}(\bv)$ over sufficiently large number of intervals should be small. In this case, we can expect the average offered service rate to be close to $\bs_{\bv^*}$. Since, $\bs_{\bv^*}$ is strictly higher than arrival rates, we can expect the average offered service rate to be strictly higher than the arrival rate. The result is formally stated next.

\begin{lemma}
\label{lem:emp_service_higher}
Consider $N(n,\epsilon):=({7\times3^5nD^2})/({\alpha\epsilon^2})$ update intervals. Then, the average of empirical service rates over these update intervals is greater than or equal to $\blam+\frac{\epsilon}{8}{\bf 1}$, i.e., $$\frac{1}{N}\sum_{l=1}^{N}\bbE\left[\hat{\bs}(l)\right] \ge \blam+\frac{\epsilon}{8}{\bf 1}.$$
\end{lemma}
\begin{IEEEproof}
See Appendix.
\end{IEEEproof}

Now, we proceed to show that the appropriate `drift' required for stability is obtained.

\subsection{Proof of Theorem \ref{thm:throughput_optimal}}
Consider the underlying network Markov chain $X(l)$ consisting of all the queues in the network, the update parameters, and the resulting rate allocation vectors at time $\tau_l$, i.e., $X(l)=(\bQ(\tau_l),\bv(\tau_l),\br(\tau_l))$ for $l\in \bbZ_+.$ It follows from the system model and the algorithm description that $X(l)$ is a time-homogenous Markov chain on an uncountable state-space $\X.$ The $\sigma$-field on $\X$ considered is the Borel $\sigma$-field associated with the product topology. For more details on dealing with general state-space Markov chains, we refer readers to \cite{Meyn-Tweedie-93}.

We consider a Lyapunov function $V:\X \rightarrow \bbR_+$ of the form, $V(\bx)=\sum_{i=1}^n (Q_i^2+v_i^2+r_i^2)$ for $\bx=(\bQ,\bv,\br)$. In order to establish positive Harris recurrence, for any $\blam$ such that $\blam + \epsilon {\bf 1} \in \T$, we use multi-step\footnote{This is a special case of the state-dependent drift criteria in \cite{Meyn-Tweedie-93}.} Lyapunov and Foster's \emph{drift} criteria to establish positive recurrence of a set of the form $V(\bx) \le \kappa$, for some $\kappa>0.$ From the assumption on the arrival processes, it follows that $V(\bx) \le \kappa$ is a closed \emph{petite} set (for definition and details see \cite{Meyn-Tweedie-93,Jiang-Shah-Shin-Walrand09}). It is well known that these two results imply positive Harris recurrence \cite{Meyn-Tweedie-93}.

Next, we obtain the required drift criteria. For simplicity, we denote $\bbE[\cdot|X(0)]$ by $\bbE[\cdot]$ in the rest of this section. Consider
\be
\bbE[Q^2_i(TN)-Q^2_i(0)]&=&\bbE[(Q_i(TN)-Q_i(0))^2\n\\
&&+2Q_i(0)(Q_i(TN)-Q_i(0))]\n\\
&\overset{(a)}{\le}&(\max(K,\bar{K})TN)^2+\n\\
&&2Q_i(0)\bbE[Q_i(TN)-Q_i(0)].\n
\ee
Here, $(a)$ follows from the fact that over unit time queue difference belong to $[-\bar{K}, K]$. Now, we look at two cases. If $Q_i(0)>\bar{K}TN$, clearly $Q_i(t)>0$ during interval $[0, TN]$ as service rate is less than or equal to $\bar{K}$. For this case, from Lemma \ref{lem:emp_service_higher}, 
\be
2Q_i(0)\bbE[Q_i(TN)-Q_i(0)] &=&2Q_i(0)T\left(\sum_{l=1}^{N}(\lambda_i-\bbE[\hat{s}_i(l)]\right)\n\\
&{\le}& -\frac{\epsilon}{4}TNQ_i(0)\n\\
&\overset{(a)}{\le}& -\frac{\epsilon}{4}TNQ_i(0)+\frac{\epsilon}{4}\bar{K}(TN)^2.\n
\ee
Here, $(a)$ is trivial, but the extra term is added to ensure that the RHS evaluates to a non-negative value for $Q_i(0)\le\bar{K}TN$. If $Q_i(0)\le\bar{K}TN$, then clearly $2Q_i(0)\bbE[Q_i(TN)-Q_i(0)] \le 2\bar{K}K(TN)^2.$ Since the bounds for each case do not evaluate to negative values for the other case, we have
\[\bbE[Q^2_i(TN)-Q^2_i(0)]\le-\frac{\epsilon}{4}TNQ_i(0)+((K+\bar{K})^2+\frac{\epsilon}{4}\bar{K})(TN)^2.\]
Since both $\bv$ and $\br$ are bounded, there exists some fixed $M(n,\epsilon)$ such that 
\[\bbE[v^2_i(TN)-v^2_i(0)]+\bbE[r^2_i(TN)-r^2_i(0)]\le M(n,\epsilon).\]

Summing up over all $i\in \calL$, we obtain
\be
\bbE[V(X(N))-V(X(0))]\le -\frac{\epsilon}{4}TN\left(\sum_{i=1}^nQ_i(0)\right)\n\\
+nM(n,\epsilon)+n\left((K+\bar{K})^2+\frac{\epsilon}{4}\bar{K}\right)(TN)^2.\n
\ee
This shows that there exists some $\kappa>0$ such that for all $\bx$ with $V(\bx) > \kappa$ there is strict negative drift. Hence, the set $V(\bx) \le \kappa$ is positive recurrent. Since $\blam + \frac{\epsilon}{2}{\bf 1}\in \R_c$, clearly $\blam + {\epsilon}{\bf 1}\in \T$. This completes the proof of Theorem \ref{thm:throughput_optimal}.
\hfill $\IEEEQED$

In summary, given any rate region for a wireless network, the (queue-length based) algorithm has $\epsilon$-optimal performance.

\section{Applications: White-space Networks}
\label{sec:application}


An important application of our algorithmic framework is in the domain of white-space networks \cite{Mitola_Maguire_99, cognitive_text_07}. White-space radios are typically required to sense the environment \cite{FCC_08}. Therefore, these radios are designed with highly accurate sensing capabilities. Even though these are primarily designed for sensing the presence of primary radios, the same capability can exploited for sensing secondary radios. In this section, we consider a networks of secondary nodes that use the same spectrum, but different from that used by primary nodes. In particular, we assume that the secondary nodes have already found spectrum that are not utilized by primary nodes. 

Since such a white-space network of secondary nodes are not centrally controlled, it is desirable to obtain simple distributed algorithms. However, the scheduling problem in these white-space networks is different from the link scheduling problem in traditional wireless networks \cite{Supratim_09}. First, the available spectrum for the operation of this network is fragmented with different propagation characteristics. Second, these secondary nodes are usually equipped with multiple radios to operate simultaneously in different bands. This is referred to as the multiple-band multiple-radio scheduling problem. Next, we describe the multiple-band multiple-radio scheduling problem in detail.

Consider the network model introduced in Section \ref{sec-system}. Define functions $s: \cL \mapsto \cN$ that maps links to source nodes, and $d: \cL \mapsto \cN$ that maps links to destination nodes. The available spectrum for the operation of this network is fragmented. The spectrum consists of $M$ bands, labeled $\cB = \{1,2,\ldots,M\}$, with bandwidths $B_1,B_2,\ldots,B_M.$ The transmission from a node to another node gets different spectral efficiencies on different bands. For a link $i$, let $c_{i,b}$ be the spectral efficiency that node $s(i)$ gets when it transmits on band $b$ to node $d(i)$. The link interference graphs are also different on different bands. Let $\cG_b=(\cL,\cE_b)$ be the link interference graph on band $b$, i.e, the transmission of link $u$ interfere with the transmission of link $v$ in band $b$ if $(u,v)\in \cE_b.$ We assume that the link interference is symmetric, i.e., if $(u,v)\in \cE_b$ then $(v,u)\in \cE_b$. These capture the frequency dependent propagation characteristics and the spatial variation of the quality of spectrum. Further, each node $j$ is equipped with $a_j$ radios. 

At time $t$, the decision whether link $i$ is operated in band $b$ is represented by binary decision variables $\sigma_{i,b}(t)$, with $1$ representing ``true'' and $0$ representing ``false''. The decision variables has to satisfy the constraints that arise from the following. \emph{(i)} Interference constraints: In every band, the set of allocated links must be non-interfering. \emph{(ii)} Radio constraints: The total number of radios at each node is limited, and these radios are half-duplex, i.e., a link requires its end nodes to dedicate one radio each for a transmission to happen. More formally, the set of constraints are:
\be
\label{eq:const1}
\sigma_{u,b}(t)+\sigma_{v,b}(t) &\le& 1, \forall (u,v)\in \cE_b, \forall b \in \cB,\\
\label{eq:const2}
\sum_{i:j\in \{s(i),d(i)\}}\sum_{b\in \cB}\sigma_{i,b}(t) &\le& a_j, \forall j \in \cN.
\ee
For a feasible schedule, the rate of flow supported on link $i$ is
\beq
r_i(t) = \sum_{b\in \cB}\sigma_{i,b}(t)c_{i,b}B_b. \n
\eeq
We denote the vector of above rates by $\br(t)$. The throughput region $\T \subseteq \bbR_+^n$ is defined as the convex hull of the set of all feasible rate vectors. Note that the queue dynamics is exactly same as described in Section \ref{sec-system}.

\subsection{Distributed Algorithm}

In this section, we present an adaptation of the developed algorithm that is throughput-optimal for multiple-band multiple-radio scheduling. For simplicity, we assume that perfect and instantaneous carrier sensing is possible on every band. The scheduling vector corresponding to link $i$ is $\bsig_i(t)=\{\sigma_{i,b}(t)\}_{b\in \cB}$. For this link, the possible states are 
\[
\{\btheta_i: \btheta_i=\{\theta_{i,b}\}_{b\in \cB}, \theta_{i,b}\in \{0,1\}, \|\btheta_i\|_0 \le \min\{a_{s(i)},a_{d(i)}\}\}.
\]
The link uses an independent exponential clock corresponding to each state with transition rate $\exp(\sum_{b\in \cB}\theta_{b}c_{i,b}B_bv_i)$ for state $\btheta$. Based on these clocks, the link obtains $\bsig_i(t)$ as follows: 
\begin{enumerate}
\item If the clock associated with a state (say $\btheta$) ticks and transitioning to that state $\bsig_i(t)=\btheta $ is feasible\footnote{This is determined using carrier sensing and radio constraints at the source and the destination of that link.}, then $\bsig_i(t)$ is changed to $\btheta$; 
\item Otherwise, $\bsig_i(t)$ remains the same. 
\end{enumerate}
The above procedure continues. The parameter $v_i$ is updated over time as a function of the queue-length $Q_i(t)$ as described in Section \ref{sec:algorithm}. This makes the algorithm completely distributed. The vector of $\{v_i\}_{i\in \cL}$ is denoted by $\bv$.

In order to establish that this algorithm is throughput-optimal, we show a correspondence between it and the rate allocation algorithm in Section \ref{sec:algorithm}. Consider a fixed $\bv$. The above algorithm forms a Markov chain on the set of feasible states. Let $S(t)$ denote the matrix formed by vectors $\{\bsig_i(t)\}_{i\in \cL}$, and $\cS$ denote the set of feasible matrices satisfying (\ref{eq:const1}) and (\ref{eq:const2}). The transition rate from state $\hat{S}=\{\hat{\btheta}_i\}_{i\in \cL}$ to state $S =\{\btheta_i\}_{i\in \cL}$ can be expressed as 
\be
q(\hat{S},S)=\left\{
\begin{array}{ll}
f(\hat{S},S), &\text{if } \sum_{i=1}^{n}\bI(\btheta_l\ne \hat{\btheta}_i)= 1,\\
0, &\text{if } \sum_{i=1}^{n}\bI(\btheta_i\ne \hat{\btheta}_i) > 1,\\
\end{array}
\right.\n
\ee
where 
\[f(\hat{S},S)=\exp\left(\sum_{i=1}^{n}\sum_{b\in \cB}\theta_{i,b}c_{i,b}B_bv_i\bI(\btheta_i\ne \hat{\btheta}_i)\right).
\] And, the diagonal elements of the rate matrix are given by $q(\hat{S},\hat{S}) = -\sum_{S \in \cS, S \ne \hat{S}} q(\hat{S},S)$ for all $\hat{S} \in \cS$.

Now, the following lemma is immediate. 
\begin{lemma}
\label{lem:finite_state_MC}
The Markov chain $(\cS,q)$ is reversible and has the stationary distribution
\beq \label{eq:st_dist}
\pi_\bv(S)&=& \frac{\exp\left(\sum_{i=1}^{n}\sum_{b\in \cB}\theta_{i,b}c_{i,b}B_bv_i\right)}{\sum_{\tilde{S} \in \cS}\exp\left(\sum_{i=1}^{n}\sum_{b\in \cB}\hat{\theta}_{i,b}c_{i,b}B_bv_i\right)}\n \\
&=& \frac{\exp\left(\br(S)\cdot \bv\right)}{\sum_{\tilde{S} \in \cS}\exp\left(\br(\hat{S})\cdot \bv\right)}.\n
\eeq
Furthermore, this Markov chain converges to this stationary distribution starting from any initial distribution.
\end{lemma}

The \emph{offered service rate} vector under the stationary distribution is 
$\bs_\bv = \sum_{S \in \cS} \pi_\bv(S) \br(S).$
Thus, we show a one-to-one correspondence to the rate allocation algorithm. As a consequence, we establish the throughput-optimality of the algorithm described in this section based on Theorem \ref{thm:throughput_optimal}.

\section{Discussion \& Simulation}
\label{sec:simulation}
\subsection{Determining Rate Feasibility}
\label{sec:feasibility}
Although our algorithm removes the control overhead associated with queue-length exchanges in the network, it still requires each link to determine rate feasibility. To elaborate, feasibility implies data-rates of other links are not impacted, i.e., other links are able to maintain their data-rates in spite of the change in the given link's data-rate. Each link can possibly change its coding and modulation strategies to ensure this. A link can determine whether a data-rate is feasible if it knows the current set of data-rates associated with other links. An important fact that makes the algorithm of practical value is that a link needs to know only data-rates associated with those links that it interferes with. Therefore, in a large network, every link needs to learn data-rates associated with few physically near-by links from control messages, for example, through ACK/NACKs when ARQ is present. We refer to the process of determining rate feasibility from the interactions of physically near-by links as ``\emph{channel measuring}''. This can be considered as a natural extension of \emph{sensing} in CSMA.

In order to further explain ``channel measuring'', we consider an example with a simplified physical-layer model. In this model, a transmitter can potentially communicate with a receiver if the receiver is within distance $d_0$. This transmitter can communicate at data-rate $r_j, 1\le j\le k,$ if there are no other transmitters within distance $d_j$ to it. We consider $r_1 \le r_2 \le \ldots \le r_k$ and $d_0\le d_1 \le \ldots \le d_k$. In this setting, for channel measuring, a transmitter needs to simply determine the distance of the nearest active transmitter. Even though we used an over simplified physical-layer model, this shows that channel measuring is a very natural technique for determining rate feasibility. Furthermore, it suggests that slightly more complicated schemes than carrier sensing may be enough to obtain significant throughput gains.

For complex physical-layer interactions, we acknowledge that channel measuring requires a well-designed physical-layer control architecture, which, by itself, is a fairly non-trivial problem. However, radios that perform complex physical-layer signaling are increasingly common and each node has access to current channel interference level, information from beacons, pilot signals and its own location. These will definitely help such radios to perform channel measuring using existing physical-layer control overhead.

\subsection{Algorithm Parameters}

In this paper, we show that the algorithm provide throughput-optimal performance for particular choice of algorithm parameters. Although this has significant theoretical value, these parameters may not be directly suitable in practice. In particular, we may have to limit the update interval length and attempt rates as large values of update interval can result in large queue-lengths, and large attempt rates can result in frequent changes in data-rates. There are certain hardware and physical-layer coding limitations on frequently changing data-rates, and frequent attempts lead to increased sensing/measuring overhead. These limitations can be easily dealt with through modified algorithm parameters.

Our approach in the paper motivates a more general class of algorithms that can be throughput-optimal for appropriate choice of parameters. We can consider the general class with update rule 
\be
v_i(\tau_{l+1}) = h\left(v_i(\tau_l), \hat{\lambda}_i(l) -\hat{s}_i(l)\right) \n
\ee
for some function $h(\cdot).$ Next, we provide a ``good'' choice of this function based on simulation results.

\begin{figure}
\centering
\epsfig{file=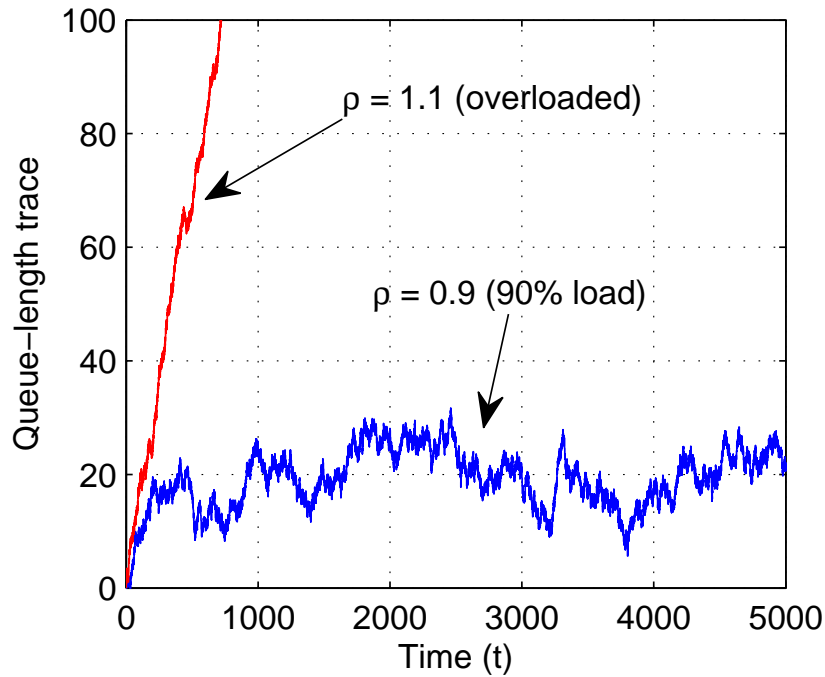,width=4.5in}
\caption{Queue-length trace from simulation}
\label{fig:MAC_simulation}
\end{figure}

\subsection{Simulation}

Consider the same Gaussian multiple access channel example with two links as before. This is shown in Figure \ref{fig:MAC_channel}. This is simply an illustrative example to show scheduling over multiple data-rate levels. Similar simulation results apply for any number of users. Let the average power constraint at the transmitters be $P = 3$ and noise variance at the receiver be $N = 1$. The information-theoretic capacity region of this channel is the pentagon shown in Figure \ref{fig:example} where $C(x)=0.5\log_2(1+x)$. The set of rate levels chosen by both transmitters are $\{0, a, b\}$ where $a=0.4$ and $b=1$. The only infeasible rate allocation pair is $[1,1]$. Consider the following arrival processes at both the transmitters. At integral times, the queues are incremented by an i.i.d. Bernoulli random variable such that the arrival rate is $\lambda = \rho \frac{a+b}{2}$, where $\rho>0$ represents the load in the system. Clearly, the network will be unstable for $\rho>1.$ 

For this system, we perform Monte-Carlo simulations with update interval $T=10$ and update rule $v_i(\tau_{l+1})=\log(1+Q(\tau_{l+1})).$ This function results in linear update near origin and prevents the rapid growth of $v_i$ with queue-length. We provide a trace of the queue-length process for $\rho=0.9$ and $\rho=1.1$ in Figure \ref{fig:MAC_simulation}. We observe that the algorithm supports $90\%$ load in the system without large increase in queue-lengths. Intuitively, this symmetric operating point is one of the difficult operating points for a distributed algorithm. More importantly, the sum-rate $\rho(a+b)$ obtained is close to the information-theoretic sum-capacity of this system. Thus, simulations show that our algorithm is of significant practice value.

\section{Conclusion}
\label{sec:concl}
Decentralized algorithms that use local sensing-based information are highly desirable in practice for wireless networks. In this paper, we develop such an algorithm that guarantees throughput-optimality. Thus, we show that efficient network algorithms can be designed that fully utilizes underlying physical-layer schemes. The algorithm is of practical value due to its decentralized nature, and due to its applications both in the newly introduced channel measurement framework, and already existing carrier sensing framework. Since this paper improves the current state-of-the-art in distributed resource allocation to account for more complex resource-sharing constraints, it has applications in other areas as well, for example, in performing resource allocation in energy networks. The algorithmic framework in this paper can be used to perform utility maximization, i.e., adaptively choose the arrival rates at the links such that a certain utility function is maximized. 

The channel measurement framework introduced in this paper motivates further research. First, we need to better understand the feasibility of channel measurement with existing and newly developed radios. This needs development of good physical-layer architectures that minimize the probability of inaccurate measurement and measurement delay. Further, we need to study the impact of imperfect channel measurement on throughput.

\section*{Acknowledgment}

The authors would like to thank S. Deb and V. Srinivasan at Bell labs, India for constructive discussions on the multiple-band multiple-radio scheduling problem.


\appendix
\section{Proof of Lemmas}
\label{sec:req_proofs}
\subsection{Optimization framework}
\subsubsection{Proof of Lemma \ref{lem:rate_stable}}
The steps involved are the following. First, we prove that, for any fixed $\blam \in \bbR_+^n$, the objective function $F(\bv,\blam)$ is \emph{strictly} concave in $\bv$. Next, we show that for any fixed $\blam \in \R_c^o$, the optimal value $\bv^*$ lies inside a compact subset of $\bbR^n$. These two statements show the existence of a unique solution that is finite. This along with certain necessary condition for optimality completes the proof. 

For notational simplicity, we denote $F(\bv,\blam)$ by $F(\bv)$ and the normalization constant or partition function by $Z(\bv):=\sum_{\br \in \R}\exp(\br \cdot \bv).$ Using calculus, it is straightforward to obtain the gradient (first-order partial derivatives) and the Hessian (second-order partial derivatives) of $F(\bv)$ in the following form:
\be
\nabla F(\bv) &=& \blam - \bbE_{\pi_\bv}[\br]\n\\
\label{eq:der3}
&=& \blam - \bs_{\bv};
\ee
\be
\label{eq:hessian}
H(F(\bv))&=& - \left(\bbE_{\pi_\bv}\left[\br \br^T\right]-\bbE_{\pi_\bv}\left[\br\right]\bbE_{\pi_\bv}\left[\br^T\right]\right).
\ee
Here, $\bs_{\bv}$ in (\ref{eq:der3}) is the offered service rate vector given by (\ref{eq:service_vector}), and $\bbE_{\pi_\bv}[\Phi]:=\sum_{\br \in \R} \pi_\bv (\br)\Phi$ for any matrix, vector or scalar $\Phi$.

In order to establish that $F(\bv)$ is \emph{strictly} concave in $\bv$, we show that the Hessian $H$ is negative definite, i.e., for any non-zero $\bet \in \R^n$, $\bet^TH\bet < 0$. Since $H$ is the negative of a covariance matrix, it is clear that $H$ is negative semi-definite, i.e., from (\ref{eq:hessian}),
\be
\bet^TH\bet&=&-\bbE_{\pi_\bv}\left[\bet^T(\br-\bbE_{\pi_\bv}\left[\br\right]) (\br-\bbE_{\pi_\bv}\left[\br\right])^T\bet \right] \n \\
\label{eq:neg_def}
&=&-\bbE_{\pi_\bv}\left[\left(\bet^T(\br-\bbE_{\pi_\bv}\left[\br\right])\right)^2 \right]\le0. 
\ee
We next prove that the Hessian $H$ is negative definite by contradiction. Consider a fixed $\bv$. Suppose that there exists $\bet \ne 0 $ such that $\bet^TH\bet=0$. Then, from (\ref{eq:neg_def}), it follows that the random variable $\bet^T(\br-\bbE_{\pi_\bv}\left[\br\right])$ is zero with probability $1$. For any fixed $\bv$, all feasible states have non-zero probability. In particular, $\pi_\bv ({\bf 0}) > 0$ and $\pi_\bv(c_i \ei) > 0$ for all $i \in \calL.$ Therefore, the random variable must evaluate to zero at $\br={\bf 0}$ and $\br=c_i \ei$, i.e., 
\[
-\bet^T\bbE_{\pi_\bv}\left[\br\right]=\eta_ic_i-\bet^T\bbE_{\pi_\bv}\left[\br\right]=0,
\]
which implies $\bet=0$. This provides a contradiction and establishes that the Hessian $H$ is negative definite.

Next, we prove that the optimal value $\bv^*$ belongs to a compact set. Let $\blam + \delta \bar{K} {\bf 1} \in \R_c$ for some $0 < \delta <1$. Note that for any $\blam \in \R_c^o$ there exists such a $\delta$. Consider a $\bv \in \bbR^n.$  Define $v_{\text{min}} = \min_i v_i$, $l = \arg\min_i v_i$, and $v_{\text{max}} = \max_i v_i$. Let 
\[
\hat{\blam} = \blam - \min(\delta \bar{K},\lambda_{\text{min}})\bI(v_{\text{min}}<0){\mathbf{e}_l}.
\]
Clearly, $\hat{\blam} + \min(\delta \bar{K},\lambda_{\text{min}}) {\bf 1} \in \R_c$, and hence, there exists a distribution $\mu$ on $\R$ such that
$
\hat{\blam} + \min(\delta \bar{K},\lambda_{\text{min}}) = \bbE_\mu[\br].
$
Since $\hat{\blam} \le \bar{K} {\bf 1}$, we have
\be
\label{eq:lamdahat}
\hat{\blam} \le \frac{\hat{\blam} + \min(\delta \bar{K},\lambda_{\text{min}})}{1+\min(\delta,\lambda_{\text{min}}/\bar{K})} = \sum_{\br \in \R} \frac{\mu(\br)\br}{1+\min(\delta,\lambda_{\text{min}}/\bar{K})} 
\ee
and
\be
\sum_{\br \in \R} \frac{\mu(\br)}{1+\min(\delta,\lambda_{\text{min}}/\bar{K})} &=& \frac{1}{1+\min(\delta,\lambda_{\text{min}}/\bar{K})} \n \\
\label{eq:muratio}
&<& 1 - \frac{\min(\delta,\lambda_{\text{min}}/\bar{K})}{2}.
\ee
From (\ref{eq:lamdahat}), (\ref{eq:muratio}) and the fact that ${\bf 0},c_i\ei \in \R$, it follows that there exists a non-negative measure $\hat{\mu}$ on $\R$ such that $\hat{\blam}  = \sum_{\br \in \R} \hat{\mu}(\br) \br$
with
$
\sum_{\br \in \R} {\hat{\mu}(\br)}= 1 -0.5\min(\delta,\lambda_{\text{min}}/\bar{K}).
$
Now, define a distribution
\be
\tilde{\mu}(\br)=
\left\{
\begin{array}{ll}
\hat{\mu}(c_l \mathbf{e}_l)+\frac{\min(\delta,\lambda_{\text{min}}/\bar{K})}{4}\bI(v_{\text{min}}<0), &\text{if } \br = c_l \mathbf{e}_l, \\
\hat{\mu}({\bf 0})+\frac{\min(\delta,\lambda_{\text{min}}/\bar{K})}{4}\left(2-\bI(v_{\text{min}}< 0)\right), &\text{if } \br = {\bf 0}, \\
\hat{\mu}(\br), &\text{otherwise}. \n
\end{array}
\right.
\ee
Define $\tilde{\blam}=\bbE_{\tilde{\mu}}[\br]$. Now, we have
\be
\tilde{\blam}= \blam - \left(1-\frac{c_l}{4\bar{K}}\right)\min(\delta \bar{K},\lambda_{\text{min}})\bI(v_{\text{min}}<0){\mathbf{e}_l}.\n
\ee
Clearly, $\blam\cdot \bv \le \tilde{\blam}\cdot \bv.$ Substituting these inequalities in (\ref{eq:opt_problem}), we obtain
\be
F(\bv)&=&\blam \cdot \bv - \log Z(\bv) \n \\
&\le&\tilde{\blam}\cdot \bv - \log Z(\bv) \n \\
&=&\sum_{\br \in \R} \tilde{\mu}(\br) \br\cdot \bv - \log Z(\bv) =\sum_{\br \in \R} \tilde{\mu}(\br)\log \frac{\exp(\br\cdot \bv)}{Z(\bv)} \n \\
&\overset{(a)}{\le}& \min\left(\tilde{\mu}(c_l \mathbf{e}_l)\log \frac{\exp(c_l \mathbf{e}_l \cdot \bv)}{Z(\bv)},\tilde{\mu}({\bf 0})\log \frac{\exp({\bf 0}\cdot \bv)}{Z(\bv)}\right) \n \\
&\overset{(b)}{\le}& \min\left(\frac{\min(\delta,\lambda_{\text{min}}/\bar{K})\bI(v_{\text{min}}<0)}{4}\log \frac{\exp(\bar{K} v_{\text{min}})}{1},\n \right.\\
\label{eq:F_bound}
&&\left. \frac{\min(\delta,\lambda_{\text{min}}/\bar{K})}{4}\log \frac{1}{\exp(\underline{K}v_{\text{max}})} \right).
\ee
Here, $(a)$ follows from $\exp(\br\cdot \bv)\le Z(\bv)$ for any $\br \in \R$, and $(b)$ follows from $\underline{K}\le c_i \le \bar{K}$ for any $i\in \calL.$ Let $\bv^* = \sup_{\bv \in \bbR^n} F(\bv)$. Then, by definition, $F(\bv^*)\ge F({\bf 0})=-\log|\R|.$ From (\ref{eq:F_bound}), we obtain the bounds
\be
\label{eq:vmax_bound}
v^*_{\text{max}}\le \frac{4\log|\R|}{\underline{K}\min(\delta,\lambda_{\text{min}}/\bar{K})}
\ee
and
\be
\label{eq:vmin_bound}
v^*_{\text{min}}\ge-\frac{4\log|\R|}{\bar{K}\min(\delta,\lambda_{\text{min}}/\bar{K})}.
\ee
Thus, there exists a unique solution which is finite. Finally, the necessary condition in (\ref{eq:der3}) for optimality completes the proof.
\hfill $\IEEEQED$

\subsubsection{Proof of Lemma \ref{lem:opt_prob_2}}
The first part of the proof follows directly from Lemma \ref{lem:rate_stable}. The second part also follows from the proof of Lemma \ref{lem:rate_stable} as explained next. In the proof, replace $\blam$ with $\blam+\frac{\epsilon}{4}{\bf 1}$ and choose $\delta = \frac{\epsilon}{4\bar{K}}.$ Now from (\ref{eq:vmax_bound}), (\ref{eq:vmin_bound}) and (\ref{eq:card_bound}), we obtain
\be
\label{eq:vinfbound2}
\|\bv^*\|_{\infty} \le \frac{4n\bar{K}\log \left\lceil {2\bar{K}}/{\epsilon}\right\rceil}{\min\left(\frac{\epsilon}{4},\lambda_{\text{min}}\right)}\frac{1}{\underline{K}}.
\ee
This follows from $\underline{K}\le \bar{K}$. If $\epsilon \le 4 \lambda_{\text{min}}$, then (\ref{eq:vinfbound2}) simplifies to (\ref{eq:vinfbound}).
\hfill $\IEEEQED$

\subsection{Mixing within update interval}
\subsubsection{Proof of Lemma \ref{lem:mix_time_bound}}
Consider the matrix $\hat{P}=\exp(P-I)$. It is fairly straightforward to verify that $\hat{P}$ corresponds the probability transmission matrix of a reversible Markov chain with the same stationary distribution $\bpi_\bv$. Now, the steps involved to complete the proof are the following. We need to obtain a lower bound on the conductance associated with $\hat{P}$ and apply Result \ref{res:conductance}. Then, we can apply Result \ref{res:mix_time} to $\hat{P}$ at $\tau = \lfloor At\rfloor$. 

From (\ref{eq:st_dist_simple}), 
$
\pi_\bv (\br) = (\exp(\br \cdot \bv))/{Z(\bv)},
$
where the partition function \[Z(\bv)=\sum_{{\br} \in \R}\exp({\br} \cdot \bv).\] From (\ref{eq:card_bound}), it is clear that $Z(\bv) \le \left\lceil {2\bar{K}}/{\epsilon}\right\rceil^n\exp(\bar{K}n\|\bv\|_{\infty}).$ 
In addition, $\exp(\br \cdot \bv) \ge \exp(-\bar{K}n \|\bv\|_{\infty})$.
Therefore, for all $\br \in \R$,
\be
\label{eq:pi_bound}
\pi_\bv (\br) \ge \frac{\exp(-2\bar{K}n \|\bv\|_{\infty})}{\left\lceil {2\bar{K}}/{\epsilon}\right\rceil^n}.
\ee

Consider two states that differ in one dimension, i.e., $\br,\hat{\br} \in \R$, $\|\br-\hat{\br}\|_0=1$, then the transition probability $\hat{P}(\br,\hat{\br})$ is lower bounded by the product of the probability that a Poisson random variable with parameter $1$ is one and $P(\br,\hat{\br})$. This follows from the fact that these two (independent) events together contribute to the transition probability $\hat{P}(\br,\hat{\br})$. Hence,
\be
\hat{P}(\br,\hat{\br}) &\ge& e^{-1} P(\br,\hat{\br})\n \\
&=&  e^{-1} \frac{f(\br,\hat{\br})}{A}\n\\
&\ge& \frac{\exp(-2\bar{K}\|\bv\|_{\infty})}{ne},\n\ee
where $f(\br,\hat{\br})$ is given by (\ref{eq:simpl_f}) and $A=n\exp(\bar{K}\|\bv\|_{\infty})$.
To lower bound conductance in (\ref{eq:conductance_def}), the following observation can be used. If both $\calS$ and $\calS^c$ are non-empty, then there is at least one state in $\calS$ and another state in $\calS^c$ that differ in one dimension alone. This follows from the fact that the state-space is connected through these one dimensional transitions alone. Applying this, we obtain
\be
\label{eq:bound_conductance}
\Phi \ge \frac{\exp(-2\bar{K}(n+1)\|\bv\|_{\infty})}{ne\left\lceil {2\bar{K}}/{\epsilon}\right\rceil^n}.
\ee
Using (\ref{eq:cheegers}), and substituting (\ref{eq:bound_conductance}), (\ref{eq:pi_bound}) in (\ref{eq:result_JSSW}), we have the required result $\|\bmu(t)-\bpi_\bv\|_{TV}\le \rho_1$ if
\[
 t = \exp\left(\Theta\left(n\|\bv\|_{\infty}+n\log\frac{1}{\epsilon}\right)\right)\log \frac{1}{\rho_1}.
\]
This completes the proof. 
\hfill $\IEEEQED$

\subsubsection{Proof of Lemma \ref{lem:error_bound}}
In the proof, we suppress $l$ in the notation, denote $\bs_{\bv}$ by $\bs$, and denote $\pi_{\bv}$ by $\pi$. From triangle inequality and linearity of expectations, we have 
\be
\bbE\left[\left\|\hat{\blam}(l) -{\blam}\right\|_1\right]+\bbE\left[\left\|\hat{\bs}(l) -{\bs_{\bv}}(l)\right\|_1\right]  \le \sum_{i=1}^n \bbE\left[|\hat{\lambda_i} -{\lambda_i} |\right] + \n \\
\label{eq:three_terms}
\sum_{i=1}^n\bbE\left[|\hat{s}_i-\bbE[\hat{s}_i]|\right] + \sum_{i=1}^n|\bbE[\hat{s}_i]-{s_i}|.
\ee
Now, we focus on $i$-th link and upper bound each of the three terms on the RHS of (\ref{eq:three_terms}) corresponding to this link separately by $\rho_2/3n$.

For bounding the first term in (\ref{eq:three_terms}), denote the arrivals over integral times as $\{\xi_k\}_{k=1}^T$. From our assumption on arrival processes, these are i.i.d. random variables with variance at most $K^2$. Hence,
\be
\bbE\left[|\hat{\lambda_i} -{\lambda_i} |\right]&\le& \left(\bbE\left[(\hat{\lambda_i} -{\lambda_i})^2 \right]\right)^{\frac{1}{2}} \n \\
&=& \left(\bbE\left[\left(\frac{1}{T}\sum_{k=1}^T{\xi_k} -{\lambda_i}\right)^2 \right]\right)^{\frac{1}{2}}\n\\
\label{eq:term1}
&\le& \frac{K}{\sqrt{T}}.
\ee

Next, we consider the expected offered service rate under distribution $\bmu(t)$, where $\bmu(t)$ denotes the distribution over $\R$ given by the algorithm at time $t$. From (\ref{eq:service_vector}), we have
\be
|\bbE_{\mu(t)}[r_i]-s_i|&=&|\bbE_{\mu(t)}[r_i]-\bbE_{\pi}[r_i]| \n \\
\label{eq:diff_dist_bound}
&\le&2\bar{K}\|\mu(t)-\pi\|_{TV}.
\ee
If we look at two times $z$ and $y$ such that $0\le z \le y$, then 
\be
\bbE[r_i(z)r_i(y)]&=& \bbE[r_i(z)\bbE[r_i(y)|r_i(z)]]\n \\
\label{eq:cond_exp_bound}
&\le&\bbE[r_i(z)]\max_{\beta\in \R_i}\bbE[r_i(y)|r_i(z)=\beta].
\ee 
We use (\ref{eq:diff_dist_bound}) and (\ref{eq:cond_exp_bound}) along with Lemma \ref{lem:mix_time_bound} to obtain bounds on the last two terms in (\ref{eq:three_terms}). Let $B(\rho_1)$ be large enough time such that it satisfies (\ref{eq:mix_time_bound}).

For the second term in (\ref{eq:three_terms}), using (\ref{eq:shat_defn}), we have
\be
&&\left(\bbE\left[|\hat{s}_i-\bbE[\hat{s}_i]|\right]\right)^2 \le \bbE\left[\left(\hat{s}_i-\bbE[\hat{s}_i]\right)^2\right]\n\\
&&= \bbE\left[\left(\hat{s}_i\right)^2\right]-\left(\bbE[\hat{s}_i]\right)^2\n\\
&&= \bbE\left[\left(\frac{1}{T}\int_{0}^{T} r_i(z) dz\right)^2\right]-\left(\bbE\left[\frac{1}{T}\int_{0}^{T} r_i(z) dz\right]\right)^2 \n\\
&&= \frac{1}{T^2}\int_{0}^{T}\int_{0}^{T}\left(\bbE\left[r_i(z)r_i(y)\right] -\bbE\left[r_i(z)\right]\bbE\left[r_i(y)\right]\right) dydz \n\\
&&= \frac{2}{T^2}\int_{0}^{T}\int_{z}^{T}\left(\bbE\left[r_i(z)r_i(y)\right] -\bbE\left[r_i(z)\right]\bbE\left[r_i(y)\right]\right) dydz\n \\
\label{eq:break_integral}
&&\le \frac{2}{T^2}\int_{0}^{T}\bbE[r_i(z)]\hat{I}dz,
\ee
where the inner integral 
\[
\hat{I}=\int_{z}^{T}\left(\max_{\beta\in \R_i}\bbE[r_i(y)|r_i(z)=\beta] -\bbE\left[r_i(y)\right]\right) dy.
\] Here, we used (\ref{eq:cond_exp_bound}). Now, from (\ref{eq:diff_dist_bound}) and Lemma \ref{lem:mix_time_bound} on mixing time, both $\max_{\beta\in \R_i}\bbE[r_i(y)|r_i(z)=\beta]$ and $\bbE\left[r_i(y)\right]$ are \emph{close} to $s_i$ by total variation $\rho_1$ each if $y\ge z+B(\rho_1)$. Formally, we bound $\hat{I}$ as follows:
\be
\hat{I} &\le&  \int_{z}^{z+{B}(\rho_1)}\bar{K}dy+\int_{z+{B}(\rho_1)}^{T}4\rho_1\bar{K} dy\n\\
\label{eq:bound_inner_int}
&\le& {B}(\rho_1)\bar{K}+4\rho_1\bar{K}T.
\ee
Substituting (\ref{eq:bound_inner_int})in (\ref{eq:break_integral}), we obtain
\be
\bbE\left[|\hat{s}_i-\bbE[\hat{s}_i]|\right] &\le& \left(\frac{2}{T^2}\int_{0}^{T}\bbE[r_i(z)]({B}(\rho_1)\bar{K}+4\rho_1\bar{K}T)dz\right)^{\frac{1}{2}} \n \\
\label{eq:term2}
&\le& \left(\frac{2}{T}\bar{K}^2{B}(\rho_1)+8\bar{K}^2\rho_1\right)^{\frac{1}{2}},
\ee
where we used $\bbE[r_i(z)]\le \bar{K}$.

For the third term, from (\ref{eq:shat_defn}) and (\ref{eq:diff_dist_bound}) and using techniques applied above, we obtain
\be
\label{eq:term3}
|\bbE[\hat{s}_i]-{s_i}|&=&\left|\frac{1}{T}\int_{0}^{T} \bbE\left[r_i(z)\right] dz-s_i\right|\n\\
&\le&\frac{\bar{K}B(\rho_1)}{T}+2\bar{K}\rho_1.
\ee
With $\rho_1=\rho_2^2/(144n^2\bar{K}^2)$ and the choice of 
\[
T=\exp\left(\Theta\left(n\|\bv\|_{\infty}+n\log\frac{1}{\epsilon}\right)\right)\frac{1}{\rho_2},
\]
it is fairly straightforward to see that RHS of (\ref{eq:term1}), (\ref{eq:term2}) and (\ref{eq:term3}) can be made smaller than $\rho_2/3n$. This completes the proof of Lemma \ref{lem:error_bound}.
\hfill $\IEEEQED$

\subsection{`Drift' over multiple intervals}
\subsubsection{Proof of Lemma \ref{lem:emp_service_higher}}
For simplicity, we denote $\bv(\tau_l)$ by $\bv_l$. Define $G(\bv):=F_{\epsilon}(\bv)-\|\bv-\bv^*\|_2^2.$ Let $[\btheta]_D$ denote component-wise $[\theta_i]_D$. This function has the following monotone property. The proof is given later in this section.
\begin{lemma}
\label{lem:monotone_fn}
Consider any $\bv \in [-D, D]^n$, $\Delta\bv \in [-1, 1]^n$. Then, $G([\bv+\Delta\bv]_D) \ge G(\bv+\Delta\bv).$ Also, $0\ge G(\bv) \ge - 7nD^2$.
\end{lemma}
Let the error term in the $l$-th time interval be 
\[
\bde_{l}=(\hat{\blam}(l) -\hat{\bs}(l))-(\blam - \bs_{\bv_l})
\]
and $\hat{\bde}_{l}=\alpha(\nabla F_{\epsilon}(\bv_l)+\bde_{l}).$ From Lemma \ref{lem:opt_prob_2}, the update equation in (\ref{eq:grad_step}) can be written as $\bv_{l+1}=\bv_l+\hat{\bde}_{l}$. We have $\nabla F_{\epsilon}(\bv_l) \in [-\bar{K}, \bar{K}]^n$, $\bde_{l}\in [-\bar{K}$ and $\bv_l,\bv^*\in[-D, D]^n$. Therefore, $\|\hat{\bde}_{l}\|_{\infty}\le \alpha(2\bar{K}+K)\le1$. From Lemma \ref{lem:monotone_fn} and Taylor's expansion, we obtain
\be
G(\bv_{l+1}) &=& G([\bv_l+\hat{\bde}_{l}]_{D})\n \\
&\ge& G(\bv_l+\hat{\bde}_{l})\n \\
&=& F_\epsilon(\bv_l+\hat{\bde}_{l}) - \|\bv_l+\hat{\bde}_{l}-\bv^*\|_2^2\n \\
&=& G(\bv_l)+\nabla F_{\epsilon}(\bv_l)\cdot\hat{\bde}_{l} +\frac{1}{2}\hat{\bde}_{l}H\hat{\bde}_{l}\n\\
\label{eq:G_break}
&&-\|\hat{\bde}_{l}\|_2^2-2(\bv_l-\bv^*)\cdot\hat{\bde}_{l},
\ee
where $H$ is the Hessian of $F_{\epsilon}(\cdot)$ evaluated at some $\tilde{\bv}$ around $\bv_l$. The elements of the matrix $H$ belong to $[-\bar{K}^2, \bar{K}^2]$, $\bde_{l}\in [-\bar{K}, K]^n$, $\nabla F_{\epsilon}(\bv_l) \in [-\bar{K}, \bar{K}]^n$ and $\bv_l,\bv^*\in[-D, D]^n$. Therefore, $\|\hat{\bde}_{l}\|_{\infty}\le \alpha(2\bar{K}+K)$. Using these, we have
\[
\frac{1}{2}\hat{\bde}_{l}H\hat{\bde}_{l}-\|\hat{\bde}_{l}\|_2^2\ge -\alpha^2c,
\]
where $c=(2\bar{K}+K)^2\left(\frac{\bar{K}^2n^2}{2}+n\right).$ Since $F_{\epsilon}(\bv)$ is concave with optimum $\bv^*$, $$F_{\epsilon}(\bv^*)\le F_{\epsilon}(\bv_l)+\nabla F_{\epsilon}(\bv_l)\cdot(\bv_l-\bv^*).$$ It follows that $\nabla F_{\epsilon}(\bv_l)\cdot(\bv_l-\bv^*)\ge0$. Applying these to (\ref{eq:G_break}), we obtain
\be
G(\bv_{l+1}) &\ge& G(\bv_l)+\alpha\|\nabla F_{\epsilon}(\bv_l)\|_2^2+\alpha\nabla F_{\epsilon}(\bv_l)\cdot\bde_{l}-\alpha^2c \n\\
&&-2 \alpha(\bv_l-\bv^*)\cdot\nabla F_{\epsilon}(\bv_l)-2\alpha(\bv_l-\bv^*)\cdot \bde_{l},\n\\
&\ge& G(\bv_l)+\alpha\|\nabla F_{\epsilon}(\bv_l)\|_2^2-\alpha\bar{K}\|\bde_{l}\|_1-\alpha^2c \n\\
&&-4\alpha D\|\bde_{l}\|_1,\n\\
&\ge& G(\bv_l)+\alpha\|\nabla F_{\epsilon}(\bv_l)\|_2^2-5\alpha D\|\bde_{l}\|_1-\alpha^2c. \n
\ee
Here, we used $\bar{K}\le D$. 

Next, performing telescopic sum and then using $G(\bv_{1}) \ge -7nD^2$ from Lemma \ref{lem:monotone_fn}, we obtain, 
\be
G(\bv_{N+1})&=&\sum_{l=1}^N(G(\bv_{l+1})-G(\bv_{l}))+G(\bv_{1})\n\\
&\ge&\alpha\sum_{l=1}^N\|\nabla F_{\epsilon}(\bv_l)\|_2^2-5\alpha D\sum_{l=1}^N\|\bde_{l}\|_1\n\\
&&-\alpha^2cN-7nD^2.\n
\ee
Since $G(\bv_{N+1}) \le 0$, and then applying (\ref{eq:error_bound_T}), we get
\be
\frac{1}{N}\sum_{l=1}^N\|\nabla F_{\epsilon}(\bv_l)\|_2^2 &\le& \frac{5D}{N}\sum_{l=1}^N\|\bde_{l}\|_1+\alpha c+\frac{7nD^2}{\alpha N}\n \\
&\le&5D\rho_2+\alpha c+\frac{7nD^2}{\alpha N}.
\ee
Applying Cauchy-Schwarz inequality, we obtain
\be
\frac{1}{N}\sum_{l=1}^{N}\bbE\left[\nabla F_{\epsilon}(\bv_l)\right]&\le& 
\frac{1}{N}\sum_{l=1}^{N}\bbE\left[\|\nabla F_{\epsilon}(\bv_l)\|_2\right]{\bf 1}\n\\
&\le& \sqrt{\frac{1}{N}\sum_{l=1}^{N}\left(\bbE\left[\|\nabla F_{\epsilon}(\bv_l)\|_2\right]\right)^2}{\bf 1}\n\\
&\le& \sqrt{\frac{1}{N}\sum_{l=1}^{N}\bbE\left[\|\nabla F_{\epsilon}(\bv_l)\|_2^2\right]}{\bf 1}\n\\
\label{eq:grad_sum_small}
&\le&\sqrt{5D\rho_2+\alpha c+\frac{7nD^2}{\alpha N}}{\bf 1}.
\ee

Next, we look at the average of the empirical service rates over $N$ update intervals. From (\ref{eq:error_bound_T}) and Lemma \ref{lem:opt_prob_2}, we obtain
\be
\frac{1}{N}\sum_{l=1}^{N}\bbE\left[\hat{\bs}(l)\right] - \blam &=& \frac{1}{N}\sum_{l=1}^{N}\bbE\left[{\bs}_{\bv_l} -\blam + \hat{\bs}(l)-{\bs}_{\bv_l}\right]\n\\
&\ge& \frac{1}{N}\sum_{l=1}^{N}\bbE\left[{\bs}_{\bv_l} -\blam\right] - \rho_2{\bf 1}\n\\
&=& \frac{1}{N}\sum_{l=1}^{N}\bbE\left[\frac{\epsilon}{4}{\bf 1}-\nabla F_{\epsilon}(\bv_l)\right] - \rho_2{\bf 1}.\n\ee
Substituting (\ref{eq:grad_sum_small}) and proceeding, we obtain 
\[
\frac{1}{N}\sum_{l=1}^{N}\bbE\left[\hat{\bs}(l)\right] - \blam \ge\left(\frac{\epsilon}{4}-\sqrt{5D\rho_2+\alpha c+\frac{7nD^2}{\alpha N}}- \rho_2\right){\bf 1}.
\]
Now, choose $\rho_2=\frac{\epsilon^2}{5\times 3^5 D}.$ Then, 
\be
\sqrt{5D\rho_2+\alpha c+\frac{7nD^2}{\alpha N}}&=&\sqrt{\frac{\epsilon^2}{3^5}+\frac{\epsilon^2}{3^5}+\frac{\epsilon^2}{3^5}}\n\\
&=&\sqrt{\frac{\epsilon^2}{3^4}}=\frac{\epsilon}{9}.
\ee
It is easy to check $\rho_2+\frac{\epsilon}{9}\le\frac{\epsilon}{8}$. This completes the proof.
\hfill $\IEEEQED$

\subsubsection{Proof of Lemma \ref{lem:monotone_fn}}
\label{sec:mon_property}
Let $\hat{\bv}=\bv+\Delta\bv$. Clearly, $\|\hat{\bv}\|_{\infty}\le D+1.$ In order to prove $G([\hat{\bv}]_D) \ge G(\hat{\bv}),$ it is sufficient to prove the following. For any dimension $i\in \calL$, $G([\hat{\bv}]_{D,i}) \ge G(\hat{\bv}),$ where $[\hat{\bv}]_{D,i}$ is defined as: the $i$-th component of $[\hat{\bv}]_{D,i}$ is same as the i-th component of $[\hat{\bv}]_D$, and all other components of $[\hat{\bv}]_{D,i}$ are same as the corresponding components of $\hat{\bv}$. It is sufficient to prove this as we can repeatedly apply $G([\hat{\bv}]_{D,i}) \ge G(\hat{\bv})$ along all dimensions to obtain $G([\hat{\bv}]_D) \ge G(\hat{\bv}).$ 

Consider any $i\in \calL$. If $\hat{v}_i \in [-D, D]$, then $G([\hat{\bv}]_{D,i}) = G(\hat{\bv})$. Therefore, the only non-trivial cases are $\hat{v}_i \in (D, D+1]$ and $\hat{v}_i \in [-(D+1), -D)$. We consider these cases separately, and apply $\left|\partial F_{\epsilon}/\partial v_i \right| \le \bar{K}$, and $\|\bv^*\|_{\infty}\le D - \bar{K}$. For $\hat{v}_i \in (D, D+1]$, we have
\be
&&G([\hat{\bv}]_{D,i}) - G(\hat{\bv}) = \n\\
&&F_{\epsilon}([\hat{\bv}]_{D,i})- F_{\epsilon}(\hat{\bv})-((D-v^*_i)^2 -(\hat{v}_i-v^*_i)^2)\n \\
&&\ge -\bar{K}(\hat{v}_i-D)+(\hat{v}_i-D)(\hat{v}_i+D-2v^*_i)\n \\
&&\ge (\hat{v}_i-D)(-\bar{K}+\hat{v}_i+D-2v^*_i) \ge 0.\n
\ee
The other case follows from similar arguments.

Since $F_{\epsilon}(\bv)\le 0$, clearly $G(\bv)\le 0$. Next, we obtain a simple lower bound on $G(\bv)$ as follows:
\be
G(\bv) &=& F_{\epsilon}(\bv)-\|\bv-\bv^*\|_2^2 \n \\
&=& (\blam+\frac{\epsilon}{4}{\bf 1})\cdot \bv -\log\left(\sum_{\tilde{\br} \in \R}\exp(\tilde{\br} \cdot \bv)\right) -\|\bv-\bv^*\|_2^2\n\\
&\ge& -\bar{K}nD - \log\left(\left\lceil {2\bar{K}}/{\epsilon}\right\rceil^n\exp(\bar{K}nD)\right)-n(2D)^2\n\\
&=& -n\left(2\bar{K}D + \log\left\lceil {2\bar{K}}/{\epsilon}\right\rceil+4D^2\right) \n \\
&\ge& -7nD^2.\n
\ee
This completes the proof. \hfill $\IEEEQED$

\end{document}

%% file: mac_channel.pstex_t
\begin{picture}(0,0)%
\includegraphics{mac_channel.pstex}%
\end{picture}%
\setlength{\unitlength}{3947sp}%
\begingroup\makeatletter\ifx\SetFigFont\undefined%
\gdef\SetFigFont#1#2#3#4#5{%
  \reset@font\fontsize{#1}{#2pt}%
  \fontfamily{#3}\fontseries{#4}\fontshape{#5}%
  \selectfont}%
\fi\endgroup%
\begin{picture}(3474,2716)(3886,-5319)
\put(6826,-4561){\makebox(0,0)[lb]{\smash{{\SetFigFont{12}{14.4}{\familydefault}{\mddefault}{\updefault}{\color[rgb]{0,0,0}$Y = X_1 + X_2 + Z$}%
}}}}
\put(3976,-2836){\makebox(0,0)[lb]{\smash{{\SetFigFont{12}{14.4}{\familydefault}{\mddefault}{\updefault}{\color[rgb]{0,0,0}$X_1$}%
}}}}
\put(3901,-5236){\makebox(0,0)[lb]{\smash{{\SetFigFont{12}{14.4}{\familydefault}{\mddefault}{\updefault}{\color[rgb]{0,0,0}$X_2$}%
}}}}
\end{picture}%